\documentclass[aps,pra,twocolumn,showpacs,superscriptaddress]{revtex4-1}
\usepackage{graphicx,color}
\usepackage{amsmath}
\usepackage{amssymb}
\usepackage{bm}

\newcommand{\Tr}{\mathop{\text{Tr}}\nolimits}
\newcommand{\ket}[1]{|{#1}\rangle}
\newcommand{\bra}[1]{\langle{#1}|}

\newcommand{\sgn}{\mathop{\text{sgn}}\nolimits}
\newcommand{\diag}{\mathop{\text{diag}}\nolimits}
\newcommand{\Var}{\mathop{\text{Var}}\nolimits}
\newcommand{\Cov}{\mathop{\text{Cov}}\nolimits}
\newcommand{\Ave}{\mathop{\mathbb{E}}\nolimits}

\definecolor{dgreen}{rgb}{0,0.5,0}

\definecolor{delete}{cmyk}{0.5,0,0,0}

\begin{document}

\title{Two-mode bosonic quantum metrology with number fluctuations}



\author{Antonella De Pasquale}
\affiliation{Scuola Normale Superiore, NEST and Istituto Nanoscienze-CNR, I-56126 Pisa, Italy}
\author{Paolo Facchi}
\affiliation{Dipartimento di Fisica and MECENAS, Universit\`a di Bari, I-70126 Bari, Italy}
\affiliation{INFN, Sezione di Bari, I-70126 Bari, Italy}
\author{Giuseppe Florio}
\affiliation{Dipartimento di Fisica and MECENAS, Universit\`a di Bari, I-70126 Bari, Italy} 
\affiliation{INFN, Sezione di Bari, I-70126 Bari, Italy}
\affiliation{Museo Storico della Fisica e Centro Studi e Ricerche
``Enrico Fermi'', Piazza del Viminale 1, I-00184 Roma, Italy}
\affiliation{Dipartimento di Meccanica, Matematica e Management, Politecnico di Bari, Via E. Orabona 4, I-70125 Bari, Italy}
\author{Vittorio Giovannetti}
\affiliation{Scuola Normale Superiore, NEST and Istituto Nanoscienze-CNR, I-56126 Pisa, Italy}
\author{Koji Matsuoka}
\affiliation{Department of Physics, Waseda University, Tokyo 169-8555, Japan}
\author{Kazuya Yuasa}
\affiliation{Department of Physics, Waseda University, Tokyo 169-8555, Japan}



\date[]{October 23, 2015}

\begin{abstract}
We search for the optimal quantum pure states  of identical bosonic particles for applications in quantum metrology, in particular in the estimation of a single parameter for the generic two-mode
interferometric setup. 
We consider the general case in which the total number of particles is fluctuating around an average $N$ with variance $\Delta N^2$. By recasting the problem in the framework of classical probability, we clarify the maximal accuracy attainable 
and show that it is always larger than the one reachable with a fixed number of particles (i.e., $\Delta N=0$). In particular, for larger fluctuations, the error in the estimation diminishes proportionally to $1/\Delta N$, below the Heisenberg-like scaling $1/N$. We also clarify the best input state, which is a ``quasi-NOON state'' for a generic setup, and for some special cases a two-mode ``Schr\"odinger-cat state'' with a vacuum component. In addition, we search for the best state within the class of pure Gaussian states with a given average $N$, which is revealed to be a product state (with no entanglement) with a squeezed vacuum in one mode and the vacuum in the other.
\end{abstract}


\maketitle

\section{Introduction}\label{sec:intro}
By making use of quantum-mechanical features, such as {quantum superposition, entanglement, or squeezing}, we are able to go beyond classical technologies.
One of such promising ideas is ``quantum metrology'' because of its possible applications \cite{ref:MetrologyScience,ref:Paris-IJQI,ref:MetrologyNaturePhoto,ref:hayashi,ref:fei,ref:dowling,ref:kobolov,ref:posit,ref:litho,ref:NOONMicroscope}.
When one wishes to estimate some quantity or parameter of a physical system, one typically tries to do it by analyzing the data collected by performing a number of independent and identical experiments, or by sending a number of independent probes to the target.
The error in the estimation scales as $1/\sqrt{N}$ (shot noise or standard quantum limit) and diminishes as the number of probes $N$ increases.
On the other hand, it has been recognized that such scaling can be beaten by quantum-mechanical effects.
In particular, the possibility of estimating phase shifts at the ``Heisenberg limit,'' with errors scaling as $1/N$, has been revealed with interferometric setups, which
exploit the possibility of employing quantum correlations in the input states of the probes
 (the optimal choice being identified  with the so-called ``NOON states'') \cite{ref:noon,ref:QuantumMetrologyVittorio}.

Analogous results hold also when the total number of probes is not exactly fixed at some  value but is allowed to fluctuate around an average number $N$.
Such situations are often found in real experiments, e.g., when measuring an optical phase difference in a two-port Mach-Zehnder interferometer \cite{CAVES,ref:HollandBurnett1993,ref:DowlingParity,PEZZE1,ref:SmerziNumberSqueezing,ref:caves1,ref:caves2,PEZZE,DOWLING}.
Also in this case, one may recognize the existence of a $1/\sqrt{N}$ scaling associated with the accuracy attainable, when 
employing classical light sources (say coherent states) as probing signals. As in the fixed-number scenario, this threshold  can be overcome 
 by properly employing probes exhibiting quantum characters (say squeezing and/or entanglement). In this case, however, the formal equivalence to the Heisenberg limit 
 appears to be not as fundamental as in  the fixed-number configuration: due to the
presence of large fluctuations in the number of probes, violations of the $1/N$ scaling of the optimal accuracy are indeed possible. 
To get realistic results, extra constraints  have to be imposed, either on the structure of 
the input signals \cite{ref:SmerziNumberSqueezing,ref:caves1,ref:caves2}, or on the amount of squeezing allowed in a single experiment \cite{CAVES}, or finally on the fluctuations of the total number of particles involved in the experiment \cite{PEZZE1,PEZZE}.

In this article, we explore the quantum metrology with bosonic particles (e.g., photons) used as probes, for the most generic two-mode interferometric setup (where the total number of probe particles is preserved).
We consider the case where  the number of bosons is not exactly fixed but can  fluctuate around an average value $N$ with a certain standard deviation $\Delta N$. 
In this  general setting, we focus on the ultimate precision limit for the estimation of a parameter $\varphi$ of the generic two-input and two-output circuit described by a scattering operator $\hat{S}_{\varphi}$ [see Fig.\ \ref{fig:BlackBox}(a)]. Under the assumption of pure input probes, we find the exact expression for the quantum Cram\'er-Rao bound $\delta\varphi\ge\delta\varphi_\text{min}$, which sets the limit to the uncertainty $\delta\varphi$ in the estimation of $\varphi$.
Specifically, one gets \cite{ref:Paris-IJQI,ref:MetrologyNaturePhoto} 
\begin{equation}  \label{PHIMIN}
\delta \varphi_\text{min} = \frac{1}{\sqrt{\nu F_Q^{(\text{max})}(\varphi)}},    \end{equation} 
where $\nu$ is the number of trials and  $F_Q^{(\text{max})} (\varphi)$ is the optimal quantum Fisher information (QFI). We show that
\begin{equation}
F_Q^{(\text{max})} (\varphi)=
\left(
|\varepsilon_+-\varepsilon_-|\sqrt{N^2+\Delta N^2}
+|\varepsilon_++\varepsilon_-| \Delta N \right)^2,
\label{eqn:h2opt0}
\end{equation}
where
$\varepsilon_{\pm}$ are constants encoding the physical properties of the generic two-port circuit (as described in Sec.\ \ref{sec:basic}).
This is the central result of the present work.
In particular, one notices that 
for an antisymmetric configuration (i.e.,  $\varepsilon_{+}=-\varepsilon_{-}$), the above expression predicts  a scaling 
$1/\sqrt{N^2+\Delta N^2}$ for $\delta \varphi_{\text{min}}$, generalizing the results obtained in Refs.\ \cite{PEZZE1} and \cite{PEZZE} for a Mach-Zehnder interferometer.
On the other hand, for a symmetric case (i.e.,  $\varepsilon_{+}=\varepsilon_{-}$), we get a $1/\Delta N$ scaling.
Such simple prototype cases are compared in 
Sec.\ \ref{sec:special}\@.
More generally, Eq.\ (\ref{eqn:h2opt0}) explicitly shows that the Heisenberg-like scaling $1/N$  for $\delta \varphi_\text{min}$ can be beaten by the presence of number fluctuations. Indeed, irrespective of the values of $\varepsilon_{\pm}$,  by exploiting large number fluctuations $\Delta N \gg N$, one can get a very small estimation error $\delta\varphi_{\text{min}} \propto 1/\Delta N\ll 1/N$.
The best input state that allows us to achieve the ultimate QFI in Eq.\ (\ref{eqn:h2opt0}) is clarified to be a ``quasi-NOON state'' (see Sec.\ \ref{sec:quasinoon}), or in some special cases a two-mode ``Schr\"odinger-cat state'' with a vacuum component {(see Sec.\ \ref{sec:schcat})}.
We also identify the best input state among pure Gaussian states and see how close we can get to the above ultimate precision by a Gaussian state, which would be much simpler to generate than the quasi-NOON state or the two-mode Schr\"odinger-cat state with vacuum (see Sec.\ \ref{sec:gaussstate}).
Remarkably, the best pure Gaussian state is a product state with no entanglement: it is simply a single-mode squeezed vacuum.

This article is organized as follows. In Sec.\ \ref{sec:basic}, we review the basic definitions of scattering operator and Fisher information in the context of quantum metrology, and set up our problem. In Sec.\ \ref{sec:classical}, we rephrase the problem of the optimization of QFI in terms of classical probability, and solve it to get the maximal QFI shown in Eq.\ (\ref{eqn:h2opt0}). 
The optimal input state that allows us to achieve the maximal QFI is {exhibited} in Sec.\ \ref{sec:OptimalState}, and some special cases are considered in Sec.\ \ref{sec:special}\@. We also solve the optimization problem within the {restricted} class of pure Gaussian states, {which are of interest in quantum optics,} in 
Sec.\ \ref{sec:OptGauss}\@.
Conclusions are finally summarized  in Sec.\ \ref{sec:conclusion}\@.

\section{Basic Setup and Framework}\label{sec:basic}
\subsection{The Model} \label{sec:scattering}
\begin{figure}
\begin{tabular}{l@{\qquad}l}
(a)&(b)\\[-3.5truemm]
\makebox(112,92){\includegraphics[width=0.22\textwidth]{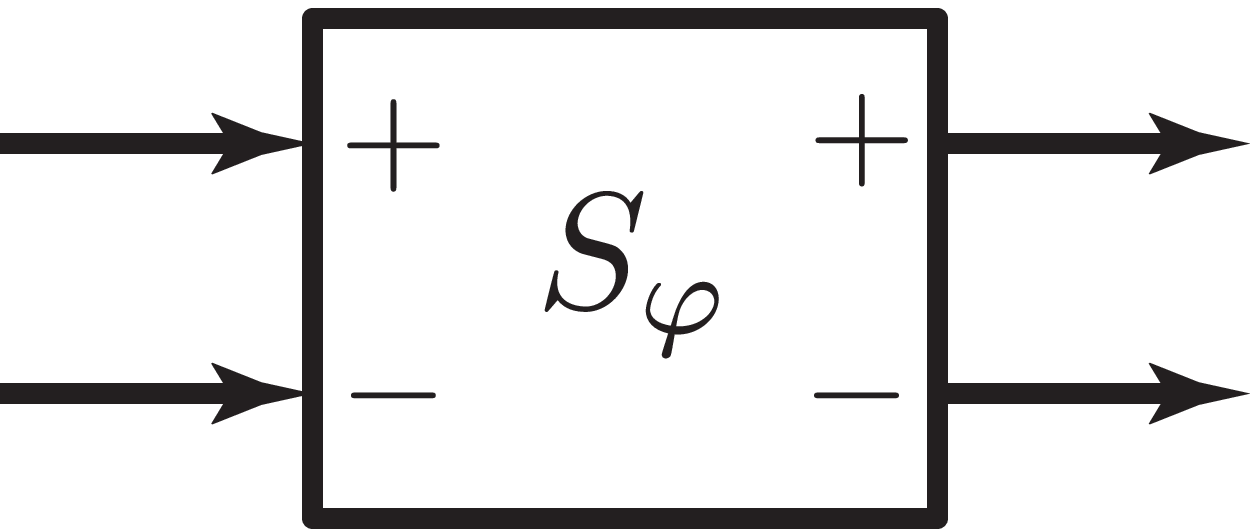}}&
\makebox(112,92){\includegraphics[width=0.22\textwidth]{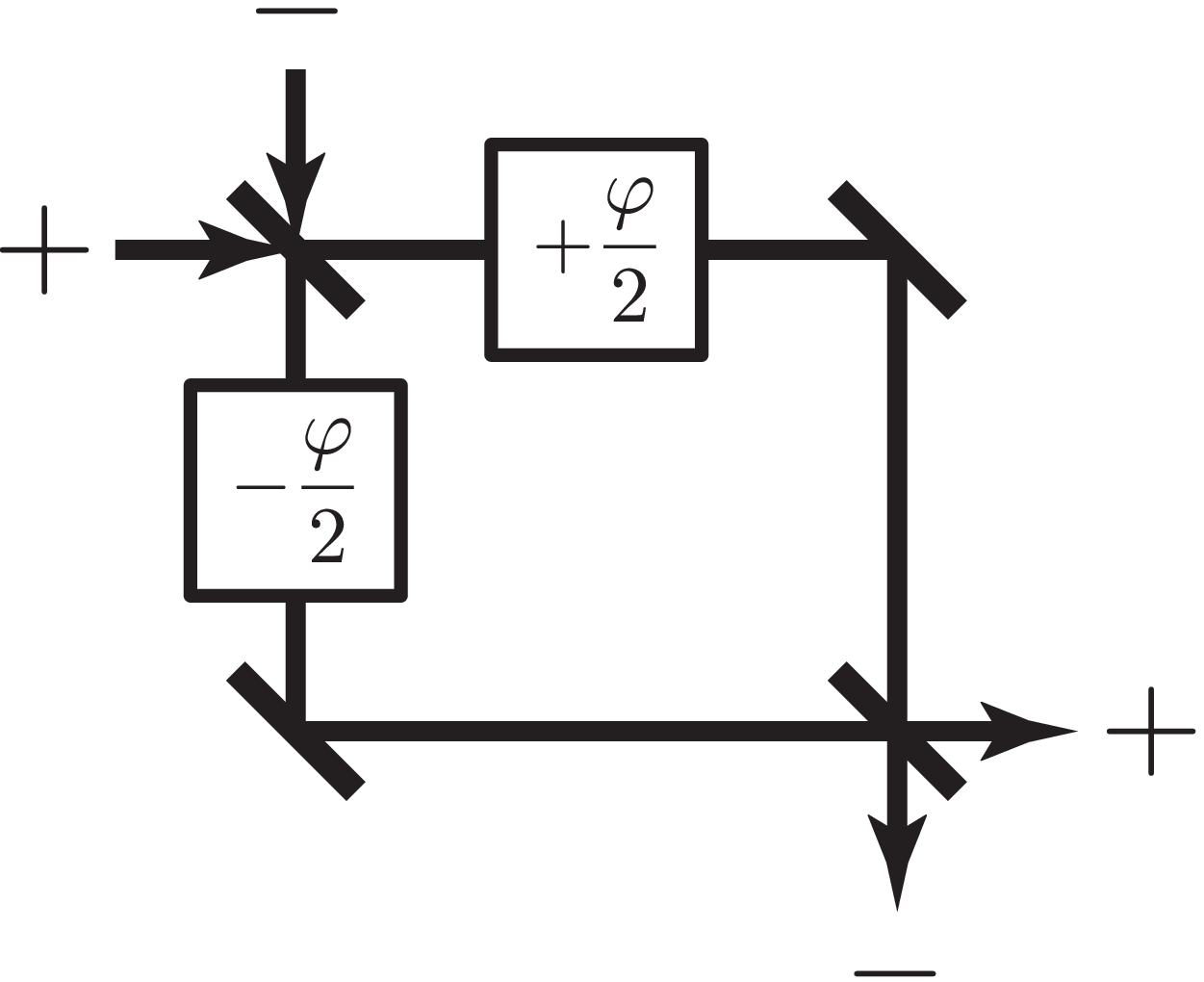}}
\end{tabular}
\caption{(a) A generic two-input and two-output linear (particle-number preserving) unitary circuit. The ports $\pm$ are associated respectively with the modes $\hat{a}_\pm$ in Eq.\ (\ref{eqn:Smatrix}). (b) Mach-Zehnder interferometer as an example of the two-input and two-output linear circuit (a). In  this special configuration, 
incoming photons go through the first balanced beam splitter described by the unitary scattering operator $\hat{U}_\text{BS}=e^{-(\pi i/4)(\hat{a}_+^\dag\hat{a}_-+\hat{a}_-^\dag\hat{a}_+)}$, acquire a relative optical phase $\varphi$ between the two paths of the interferometer by $\hat{V}_\varphi=e^{-i(\varphi/2)(\hat{a}_+^\dag\hat{a}_+-\hat{a}_-^\dag\hat{a}_-)}$, and pass the last balanced beam splitter $\hat{U}_\text{BS}^\dag$. The overall unitary transformation is therefore given by
$\hat{S}_\varphi=\hat{U}_\text{BS}^\dag\hat{V}_\varphi\hat{U}_\text{BS}=e^{(\varphi/2)(\hat{a}_-^\dag\hat{a}_+-\hat{a}_+^\dag\hat{a}_-)}$,
corresponding  to setting the parameters in Eq.\ (\ref{eqn:TRpara}) to 
$\beta=\varphi/2$, $\chi=\tau=0$, and $\rho= -\pi/2$, which yield the matrix elements in Eq.\ (\ref{eqn:Smatrix}) equal to
$T_\pm=\cos(\varphi/2)$ and $R_\pm=\pm\sin(\varphi/2)$.
The generator $\hat{H}_\varphi$ in Eq.\ (\ref{eqn:Hg2}) is instead given by 
$\hat{H}_\varphi=(i/2)(\hat{a}_-^\dag\hat{a}_+-\hat{a}_+^\dag\hat{a}_-)$, corresponding to setting the functions in Eq.\ (\ref{eqn:AB}) to  $A_\pm=0$ and $B=i/2$.
In the diagonalized form (\ref{eqn:Hcd}), we have 
 $\varepsilon_\pm=\pm1/2$.}
\label{fig:BlackBox}
\end{figure}

We consider a generic two-input and two-output linear (particle-number preserving) unitary circuit [Fig.\ \ref{fig:BlackBox}(a)].
Our problem is to estimate a single parameter $\varphi$ of the circuit, by injecting  bosonic particles {into} the circuit and observing its output.
A typical example is the Mach-Zehnder interferometer {used} to measure an optical phase shift $\varphi$ by injecting photons into the input ports [Fig.\ \ref{fig:BlackBox}(b)].
The following analysis however is not restricted to such a specific setup, and the parameter $\varphi$ can be something more general. 
In  particular, we describe the action of the circuit on the input state by a scattering operator  $\hat{S}_\varphi$ preserving the total number of particles.
It  induces the canonical transformation
\begin{equation}
\hat{S}_\varphi
\begin{pmatrix}
\medskip
\hat{a}_+^\dag\\
\hat{a}_-^\dag
\end{pmatrix}
\hat{S}_\varphi^\dag
=\begin{pmatrix}
\medskip
T_+&R_+\\
R_-&T_-
\end{pmatrix}
\begin{pmatrix}
\medskip
\hat{a}_+^\dag\\
\hat{a}_-^\dag
\end{pmatrix},
\label{eqn:Smatrix}
\end{equation}
where $\hat{a}_\pm^\dag$ are the creation operators for bosons incoming to and outgoing from the ports of the circuit labeled ``$\pm$'', 
and where   $T_\pm$ and $R_\pm$ 
 are complex-valued functions of $\varphi$ which define a $2\times 2$ unitary matrix by fulfilling the constraints
\begin{equation}
\begin{cases}
\medskip
\displaystyle
|T_\pm|^2+|R_\pm|^2=|T_\pm|^2+|R_\mp|^2=1,\\
\displaystyle
T_\pm^*R_\mp+R_\pm^*T_\mp=T_\pm^*R_\pm+R_\mp^*T_\mp=0.
\end{cases}
\label{eqn:PR}
\end{equation}
When feeding the device with a two-mode input state $|\psi_0\rangle$ (this is the initial state of the probing signal), it outputs the state
\begin{equation} \label{theout}
|\psi_\varphi\rangle = \hat{S}_\varphi |\psi_0\rangle ,
\end{equation}
which is the one we can monitor in order to recover the value of the parameter $\varphi$.

\subsection{Normal Modes} \label{NORMAL}
The conditions in Eq.\ (\ref{eqn:PR}) lead to the parameterization of the matrix elements
\begin{equation}
T_\pm=e^{-i\chi}e^{\mp i\tau}\cos\beta,
\quad
R_\pm=-ie^{-i\chi}e^{\mp i\rho }\sin\beta,
\label{eqn:TRpara}
\end{equation}
with $\beta$, $\chi$, $\tau$, and $\rho$ being arbitrary real-valued functions of $\varphi$.
Accordingly,  $\hat{S}_\varphi$ can be expressed as
\begin{align}
\hat{S}_\varphi
={}&e^{-i\chi(\hat{a}_+^\dag \hat{a}_++\hat{a}_-^\dag \hat{a}_-)}
e^{-i\tau(\hat{a}_+^\dag \hat{a}_+-\hat{a}_-^\dag \hat{a}_-)}
\nonumber\\
&
{}\times
e^{-i\beta(e^{-i(\tau+\rho)}\hat{a}_-^\dag\hat{a}_++e^{i(\tau+\rho)}\hat{a}_+^\dag\hat{a}_-)}.
\label{eqn:Sexp}
\end{align}
We then introduce its generator by 
\begin{align}
\hat{H}_\varphi
&=i\hat{S}_\varphi^\dag\frac{\partial \hat{S}_\varphi}{\partial\varphi}
\nonumber\\
&=
A_+\hat{a}_+^\dag \hat{a}_++A_-\hat{a}_-^\dag \hat{a}_-
+B\hat{a}_-^\dag\hat{a}_+
+B^*\hat{a}_+^\dag\hat{a}_-,
\label{eqn:Hg2}
\end{align}
where $A_\pm$ and $B$ are  the following implicit functions of $\varphi$:
\begin{equation}
\begin{cases}
\medskip
\displaystyle
A_\pm
=\frac{\partial\chi}{\partial\varphi}
\pm\frac{1}{2}\left(
\frac{\partial(\tau+\rho)}{\partial\varphi}
+\frac{\partial(\tau-\rho)}{\partial\varphi}\cos2\beta\right),
\\
\displaystyle
B
=\left(
\frac{\partial\beta}{\partial\varphi}
+\frac{i}{2}\frac{\partial(\tau-\rho)}{\partial\varphi}\sin2\beta
\right)
e^{-i(\tau+\rho)}.
\end{cases}
\label{eqn:AB}
\end{equation}
The generator $\hat{H}_\varphi$ defined as in Eq.\ (\ref{eqn:Hg2}) plays an important role in evaluating the optimal precision of the estimation of the parameter $\varphi$ later.
Since $\hat{H}_\varphi$ is Hermitian, by explicit diagonalization we can write
\begin{equation}
\hat{H}_\varphi
=\varepsilon_+\hat{c}_+^\dag\hat{c}_+
+\varepsilon_-\hat{c}_-^\dag\hat{c}_-,
\label{eqn:Hcd}
\end{equation}
where 
\begin{align}
\varepsilon_\pm
={}&\frac{1}{2}\sgn(A_+ + A_-)
\nonumber\\
&{}\times
\left(
{|A_++A_-|}\pm\sqrt{(A_+-A_-)^2+4|B|^2}
\right),
\label{eqn:Epsilon}
\end{align}
with $\sgn x = 1$ for $x\geq 0$ and $-1$ otherwise, and where
\begin{equation}
\hat{c}_{\pm}
=\frac{
B\hat{a}_+
+(\varepsilon_\pm - A_+) \hat{a}_-
}{\sqrt{(\varepsilon_\pm - A_+)^2 + |B|^2}}
\end{equation}
are the normal modes satisfying the canonical commutation relations. 
They are related to $\hat{a}_\pm$ by a unitary transformation preserving the total number of probe particles,
\begin{equation}
\hat{N}
=\hat{a}_+^\dag\hat{a}_++\hat{a}_-^\dag\hat{a}_-
=\hat{c}_+^\dag\hat{c}_++\hat{c}_-^\dag\hat{c}_-.
\label{enne} 
\end{equation} 
Notice that $\hat{H}_\varphi$ and $\hat{N}$ commute and admit as simultaneous eigenvectors the Fock states $|m,n\rangle$ of the normal modes $\hat{c}_+$ and $\hat{c}_-$ (with $m$ and $n$ being non-negative
integers), the associated spectra {being respectively} 
\begin{equation} \label{eigen} 
E_{m,n} = \varepsilon_+ m + \varepsilon_- n,    \qquad 
 N_{m,n} = m+n . 
\end{equation}
Notice also that, for our convenience, we ordered the eigenvalues $\varepsilon_{\pm}$ (and the corresponding normal modes $\hat{c}_{\pm}$) such that
\begin{equation}
\label{eq:ordering}
\varepsilon_+^2 \geq \varepsilon_-^2.
\end{equation}

{\subsection{Fisher Information}\label{sec:fisher}}
The simplest strategy for estimating the parameter $\varphi$ of the two-mode linear circuit in Fig.\ \ref{fig:BlackBox}(a) is to inject a single probe particle {into} port $+$ and see whether it is output from port $+$ or from port $-$.
We repeat this experiment many times (but a finite number of trials $\nu$), collect {the} data, and evaluate the probabilities 
$P_+=P(+|\varphi)=|T_+|^2=\cos^2\beta$ and $P_-=P(-|\varphi)=|R_+|^2=\sin^2\beta$
 for the respective possible outcomes.
By comparing these probabilities with their theoretical predictions, the parameter $\varphi$ is estimated.
The ultimate precision of this estimation can be evaluated on the basis of the Cram\'er-Rao inequality {\cite{ref:kay,ref:lehmann,ref:BraunsteinCave1994,ref:BraunsteinCave1996AnnPhys,ref:Paris-IJQI,ref:MetrologyNaturePhoto}}: the uncertainty $\delta\varphi$ in the estimation of $\varphi$ is bounded as
\begin{equation}
\delta\varphi\ge\frac{1}{\sqrt{\nu F(\varphi)}},
\end{equation}
where 
\begin{equation}
F(\varphi)=\sum_{s=\pm}P(s|\varphi)\left(\frac{\partial}{\partial\varphi}\ln P(s|\varphi)\right)^2 = \frac{1}{P_+P_-}\left(\frac{\partial P_+}{\partial\varphi}\right)^2 \label{eqn:FI}
\end{equation}
is the Fisher information (FI) of the procedure. 
Different detection strategies, i.e., not simply checking whether the probe particle comes out of \textit{either} port $+$ \textit{or} port $-$ but, for instance, checking whether the output state of the probe particle is in a superposition state of the outputs from port $+$ \textit{and} port $-$, might provide us with better estimation.
The maximum FI attainable by means of an optimal measurement is known to be expressed by the so-called QFI \cite{ref:BraunsteinCave1994,ref:BraunsteinCave1996AnnPhys,ref:Paris-IJQI,ref:MetrologyNaturePhoto,ref:Jarzyna}, which for the  pure output state $\ket{\psi_\varphi}$ in Eq.\ (\ref{theout}) 
 can be computed as the variance  of the generator $\hat{H}_\varphi$ in Eq.\ (\ref{eqn:Hg2})
 evaluated in the input state $|\psi_0\rangle$, i.e., 
 \begin{equation} 
(\Delta\hat{H}_\varphi)_{\psi_0}^2 =  \bra{\psi_0}\hat{H}_\varphi^2\ket{\psi_0}
-\bra{\psi_0}\hat{H}_\varphi\ket{\psi_0}^2 \label{variance0}.
\end{equation} 
Specifically we have
\begin{equation}
F_Q(\varphi)
=4
\frac{\partial\bra{\psi_\varphi}}{\partial\varphi}
(1-\ket{\psi_\varphi}\bra{\psi_\varphi})
\frac{\partial\ket{\psi_\varphi}}{\partial\varphi}
 = 4(\Delta\hat{H}_\varphi)_{\psi_0}^2  .
\label{eqn:FQdef}
\end{equation}
For the above-mentioned  strategy, where we input a particle from port $+$, one has $|\psi_0\rangle= \hat{a}_+^\dag\ket{0}$
and hence 
\begin{equation}
F_Q(\varphi)=\frac{1}{P_+P_-}\left(\frac{\partial P_+}{\partial\varphi}\right)^2
+4P_+P_-\left(
\frac{\partial(\tau-\rho)}{\partial\varphi}
\right)^2,
\label{eqn:QFI1}
\end{equation}
which is clearly larger than or equal to $F(\varphi)$ in Eq.\ (\ref{eqn:FI}). 
What is important is the fact that  formula (\ref{eqn:FQdef}) can be optimized by tuning the input state $\ket{\psi_0}$  within the set $\mathfrak{C}$ of states allowed by the constraints we impose on the problem \cite{ref:QuantumMetrologyVittorio,ref:MetrologyNaturePhoto}, i.e., 
\begin{equation}\label{QFIMAX}
F_Q^{(\text{max})}(\varphi) = \max_{|\psi_0\rangle \in \mathfrak{C}} F_Q(\varphi).
\end{equation} 
  Via  the quantum 
Cram\'er-Rao inequality {\cite{ref:kay,ref:lehmann,ref:BraunsteinCave1994,ref:BraunsteinCave1996AnnPhys,ref:Paris-IJQI,ref:MetrologyNaturePhoto} this 
provides the minimal uncertainty in the estimation of $\varphi$
through Eq.\ (\ref{PHIMIN}).

Determining $F_Q^{(\text{max})}(\varphi)$ for a generic two-mode linear circuit is the goal of the present work. 
In particular, we study the scenario where constraints are imposed on the total number of particles involved in the procedure.
Specifically, given $N$ and $\Delta N$ positive constants,  we address the case where the set  $\mathfrak{C}$ identifies all pure states \cite{NOTE1} 
which have an average  number of particles equal to $N$ and a variance equal to $\Delta N^2$, i.e., 
\begin{align} 
\mathfrak{C}
&=\mathfrak{C}_{N,\Delta N}
\nonumber\\
&= \{ |\psi_0\rangle \; |\;  \langle \psi_0 | {\hat{N}} |\psi_0\rangle = N  \ \text{and}  \ (\Delta\hat{N})_{\psi_0}^2 = \Delta N^2\},
\label{constraint} 
\end{align} 
where $(\Delta\hat{N})_{\psi_0}^2 =  \bra{\psi_0}\hat{N}^2\ket{\psi_0}
-\bra{\psi_0}\hat{N}\ket{\psi_0}^2$.
The case $\mathfrak{C}_{N,\Delta N=0}$ corresponds to the fixed-number scenario where exactly $N$ particles enter the circuit.

\section{Optimization of QFI as a Problem of Classical Probability}\label{sec:classical}
We notice that the generator $\hat{H}_\varphi$ in Eq.\ (\ref{eqn:Hcd}) and the particle number operator $\hat{N}$ in Eq.\ (\ref{enne}) are both composed of the sum of commuting observables, namely, the number operators $\hat{c}_\pm^\dag\hat{c}_\pm$.
Therefore, the problem of the optimization (\ref{QFIMAX}) of the QFI under the constraints (\ref{constraint})  can be analyzed in terms of classical probabilities, when considered in the basis $\{\ket{m,n}\}$ diagonalizing $\hat{c}_+^\dag\hat{c}_+$ and $\hat{c}_-^\dag\hat{c}_-$ simultaneously, with eigenvalues $m$ and $n$, respectively.
Indeed, for a generic  pure input state
\begin{equation} \label{inputstate}
\ket{\psi_0}=\sum_{m,n\ge0}\chi_{m,n}\ket{m,n},
\end{equation}
we will see that the only relevant parameters for the problem are the classical probabilities
\begin{equation}
P_{m,n}={|\chi_{m,n}|}^2,\qquad\sum_{m,n\ge0}P_{m,n}=1,
\label{eq:clprob}
\end{equation}
the phases of the amplitudes $\chi_{m,n}$ being completely irrelevant.
To see this explicitly, for a generic function $F_{m,n}$ of the integer variables $m$ and $n$, define   
\begin{equation} 
\Ave[ F_{m,n}] = \sum_{m,n\ge0} P_{m,n} F_{m,n}\label{fff},
\end{equation} 
which is the average of $F_{m,n}$ with respect to the probability distribution $P_{m,n}$. 
Then it is simple to verify that the variance (\ref{variance0}) of $\hat{H}_\varphi$ on $|\psi_0\rangle$
can be written as 
\begin{align} 
(\Delta\hat{H}_\varphi)_{\psi_0}^2
&=\Var[ E_{m,n}] 
\nonumber\\
&=\varepsilon_+^2\Var[m] +\varepsilon_-^2\Var[n]
+2\varepsilon_+\varepsilon_-\Cov[m,n],\label{variance}
\end{align}
where $E_{m,n}$ are the eigenvalues (\ref{eigen}) of $\hat{H}_\varphi$, while 
 \begin{equation}
\begin{cases}
\medskip
\displaystyle
 \Var[F_{m,n}]=\Ave[F_{m,n}^2]-\Ave[F_{m,n}]^2,\\
\displaystyle
 \Cov[F_{m,n},G_{m,n}]=\Ave[F_{m,n}G_{m,n}]-\Ave[F_{m,n}]\Ave[G_{m,n}]
\end{cases}
\end{equation}
are the variance and the covariance on $P_{m,n}$, respectively.
Similarly, the constraints (\ref{constraint}) are expressed as
\begin{align} 
\Ave[m+n] &= N \label{eqn:Ave},\\
\Var[m+n] &= \Delta N^2 \label{constraint1},
\end{align} 
which have to be obeyed by the probability distribution $P_{m,n}$.
In this way, the variance of $\hat{H}_\varphi$ in Eq.\ (\ref{variance}) (and hence the QFI) and the constraints (\ref{eqn:Ave}) and (\ref{constraint1}) are all independent of the phases of the amplitudes $\chi_{m,n}$, and the phases are irrelevant to the optimization of the QFI\@.
The optimization of the QFI is thus reduced to the optimization of the classical probabilities (\ref{eq:clprob}).

\subsection{Fixed Number of Probes}\label{sec:fixedPartNumb}
Consider first the simple case where we look for the maximum of the variance (\ref{variance}), namely, of the QFI {in Eq.\ (\ref{eqn:FQdef})}, under the constraint {that} the total number of impinging particles is fixed and equal to some integer value $N$ with no fluctuation (i.e., $\mathfrak{C}=\mathfrak{C}_{N,\Delta N=0}$). 
This implies  that the possible input states $|\psi_0\rangle$ of the problem have  to be the eigenstates {of the total number operator $\hat{N}$ with eigenvalue $N$}, i.e.,  {states} of the form (\ref{inputstate}) characterized by amplitudes $\chi_{m,n}$ 
different from 0 only for values of $m$ and $n$ fulfilling the condition $m+n=N$.
Obviously, this  forces the two variables to be linearly dependent. In particular, this implies that the joint probabilities $P_{m,n}$ should collapse to the single variable distribution $P_n = P_{N-n,n}$.
We seek for a joint probability distribution $P_{m,n}$ that maximizes the variance (\ref{variance}) under such a constraint. 
Under this condition Eq.\ (\ref{variance}) becomes
\begin{equation}
(\Delta\hat{H}_\varphi )^2_{\psi_0} =\Var[N\varepsilon_+-(\varepsilon_+-\varepsilon_-)n] = (\varepsilon_+-\varepsilon_-)^2 \Var[n],
\end{equation}
so that the problem reduces to finding the probability distribution
{$P=\{P_n\}_{n=0,\ldots, N}$} that maximizes the variance $\Var[n]$ of a {positive} integer random variable $n$ with values in $\{0,\dots, N\}$. We get
\begin{equation}
\max_P\text{Var}[n] = \max_P \frac{1}{2} \sum_{m,n=0}^{N}P_m P_n (m-n)^2 = \frac{1}{4}N^2
\end{equation}
at the unique point
\begin{equation}\label{eq:solnoon}
P_n=\frac{1}{2}\delta_{n,0}+\frac{1}{2}\delta_{n,N}.
  \end{equation}
The optimal state that maximizes the variance of the generator $\hat{H}_\varphi$ is now clear: it is the superposition of the state belonging to the maximum eigenvalue $\ket{N,0}$ and the state belonging to the minimum eigenvalue $\ket{0,N}$, with an arbitrary phase $\phi$,
\begin{equation}
\ket{\psi_0}=\ket{\text{NOON}} = \frac{1}{\sqrt{2}}(\ket{N,0} + e^{i\phi}\ket{0,N}).
\label{eqn:NOON}
\end{equation}
This is a so-called NOON state \cite{ref:MetrologyNaturePhoto,ref:QuantumMetrologyVittorio,ref:BenattiBraunIdPRA}.
{With such an input state} the QFI (\ref{eqn:FQdef}) is maximal and reads 
\begin{equation}
F_Q^{(\text{max})} (\varphi)
=N^2(\varepsilon_+-\varepsilon_-)^2
=N^2[(A_+-A_-)^2+4|B|^2] ,
\label{eqn:QFINOON}
\end{equation}
which corresponds to a minimal uncertainty $\delta \varphi_\text{min}$  in Eq.\ (\ref{PHIMIN})  exhibiting the typical $1/N$ Heisenberg scaling 
(for comparison, observe that the QFI associated with an arbitrary Fock state $|N-n, n\rangle$ is always null). 
Notice also that in the symmetric case, $\varepsilon_+ = \varepsilon_-$, the maximal QFI (\ref{eqn:QFINOON}) vanishes, implying that in this circumstance the parameter $\varphi$ cannot be recovered using
states with a fixed number of particles, the minimal uncertainty (\ref{PHIMIN}) being unbounded.

\subsection{General Setting}\label{sec:fixedVariance}
Consider next the case with nonvanishing $\Delta N$. In other words,
we do not fix the total number of particles $m+n$ as in the previous section, but let it fluctuate around its average $N$ with a variance $\Delta N^2$.
Our task is to optimize the QFI, i.e., to optimize $(\Delta\hat{H}_\varphi)_{\psi_0}^2$ in Eq.\ (\ref{variance}), under the constraints (\ref{eqn:Ave}) and (\ref{constraint1}) on the total number of particles.

We first notice that for the symmetric case with $\varepsilon_+=\varepsilon_-=\varepsilon$ the eigenvalues of the generator $\hat{H}_\varphi$ are given by $E_{m,n} = \varepsilon (m+n)$, which are proportional to the eigenvalues $m+n$ of the total number of particles $\hat{N}$. Therefore, the maximization of $(\Delta\hat{H}_\varphi)_{\psi_0}^2$ under the constraint (\ref{constraint1}) is trivial, because 
$(\Delta\hat{H}_\varphi)_{\psi_0}^2=\varepsilon^2 \Delta N^2$ for all $|\psi_0\rangle \in \mathfrak{C}_{N,\Delta N}$, 
yielding 
\begin{equation}\label{QFIMAX11}
F_Q^{(\text{max})}(\varphi) = 4 \varepsilon^2 \Delta N^2. 
\end{equation} 
Let us hence assume that $\varepsilon_+\neq\varepsilon_-$.
This is always the case if $B\neq0$ [see Eq.\ (\ref{eqn:Epsilon})].

To simplify the notation we set
\begin{align}
 x:=\Var[m], \quad
 y&:=\Var[n],\quad
 z:=\Cov[m,n],
\\
h^2&:=(\Delta\hat{H}_\varphi)_{\psi_0}^2.
\label{eq:small_c_d_h}
\end{align}
In this notation, Eqs.\ (\ref{variance}) and  (\ref{constraint1}) read
\begin{gather}
h^2=\varepsilon_+^2 x + \varepsilon_-^2 y + 2\varepsilon_+\varepsilon_- z, \label{eq:hmax} \\
 x + y+2 z= \Delta N^2,  \label{eq:var}
\end{gather}
with $x,y\ge 0$. 
Using the constraint (\ref{eq:var}), $z$ is removed from the formula for $h^2$ in Eq.\ (\ref{eq:hmax}) to yield
\begin{equation}
h^2
=(\varepsilon_+-\varepsilon_-)^2\frac{x+y}{2}+(\varepsilon_+^2-\varepsilon_-^2)\frac{x-y}{2}+\varepsilon_+\varepsilon_- \Delta N^2,
\label{eq:h2}
\end{equation}
which is a function of $x$ and $y$.
Our problem is therefore to maximize $h^2$ in Eq.\ (\ref{eq:h2}) within the region on the $xy$-plane allowed under the constraints (\ref{eqn:Ave}) and (\ref{eq:var}).

The region on the $xy$-plane is limited as follows.
First, the covariance $z=\Cov[m,n]$ is bounded by the Cauchy-Schwarz inequality $|{\Cov[m,n]}|\leq \sqrt{\Var[m]\Var[n]}$, i.e.,
\begin{equation}
|z|\le\sqrt{xy}.
\label{eqn:CS2}
\end{equation}
Combined with Eq.\ (\ref{eq:var}) it yields the bounds on $x$ and $y$,
\begin{equation}
 -2 \sqrt{xy}\le x+y - \Delta N^2 \le 2 \sqrt{xy}.
 \label{eqn:BoundXY1}
\end{equation}
Its boundary is given by a parabola on the $xy$-plane,
\begin{equation}\label{eq:constraint2}
(x-y)^2- 2\Delta N^2 (x+y)+ \Delta N^4=0,
\end{equation}
which is symmetric with respect to the line $x=y$ and {tangent to} the $x$ and $y$ axes at $(\Delta N^2,0)$ and $(0,\Delta N^2)$, respectively. 
See Fig.\ \ref{fig:xyplane}(a).
\begin{figure}[t]
\begin{tabular}{l@{}l}
(a)&(b)\\[-3.5truemm]
\hphantom{(a)}\ \includegraphics[height=0.21\textwidth]{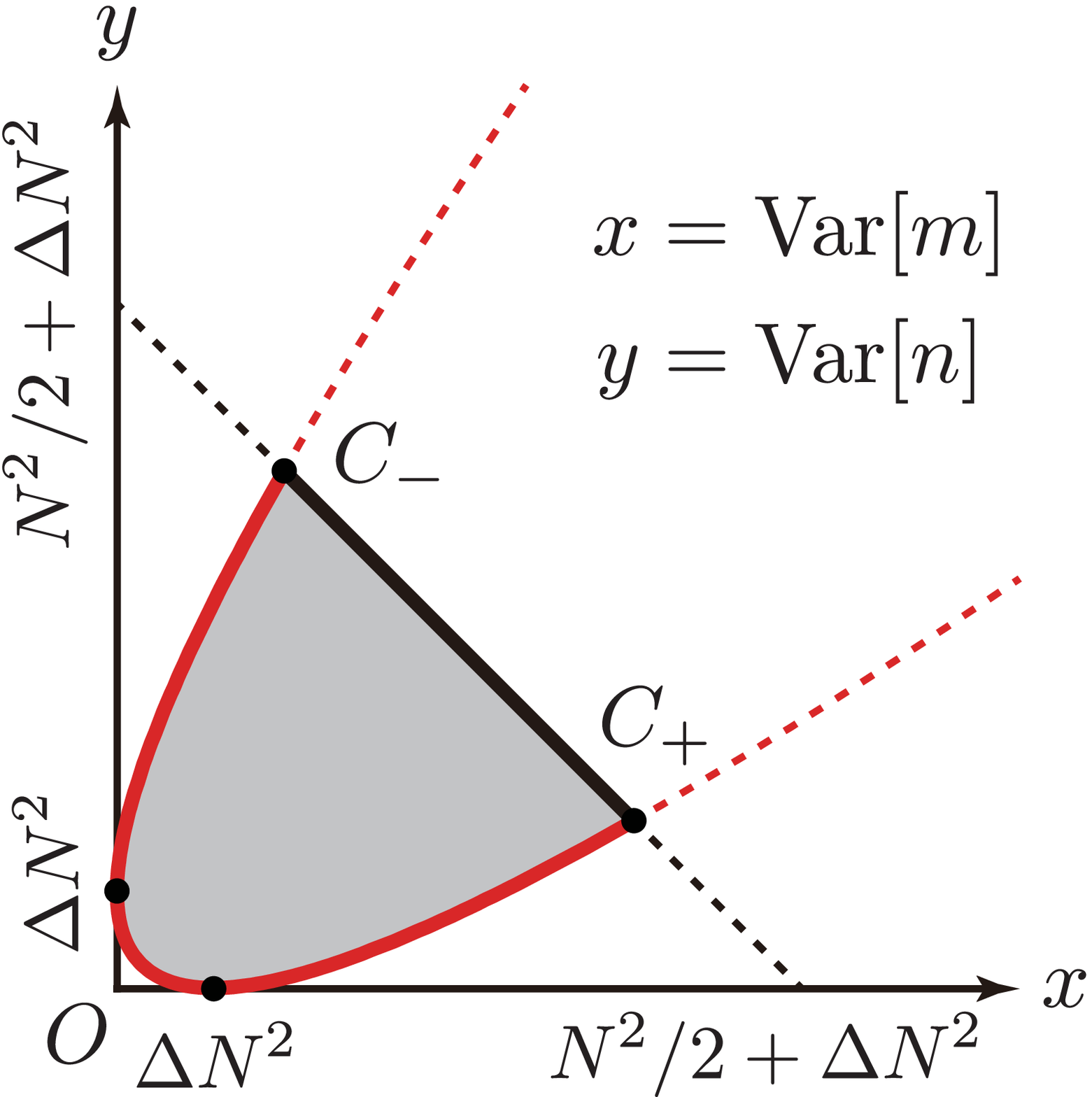}&
\hphantom{(b)}\includegraphics[height=0.21\textwidth]{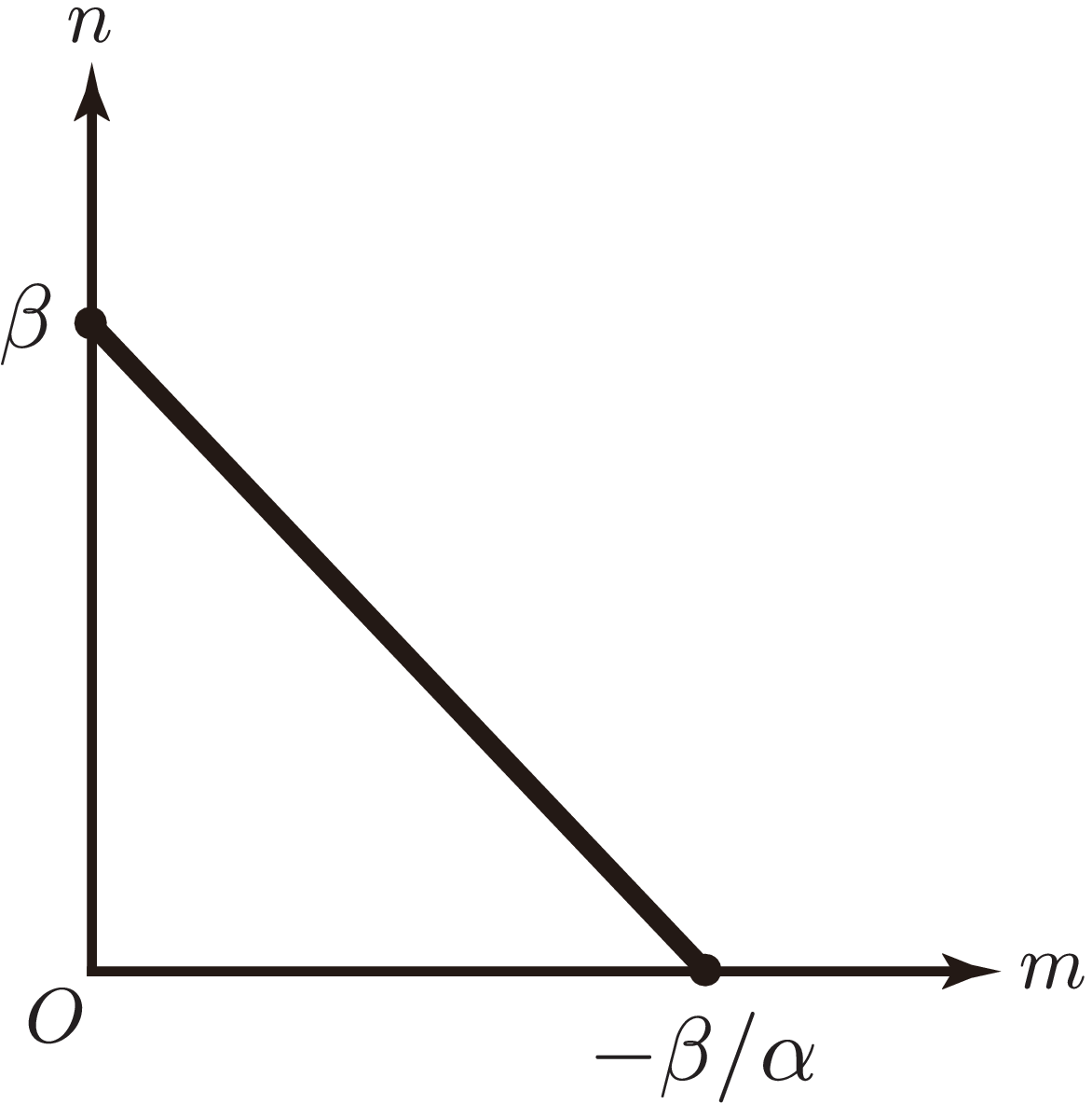}
\end{tabular}
\caption{(a) The domain on the variance $xy$-plane (shaded region) where maximization of the function $h^2 =(\Delta\hat{H}_\varphi)_{\psi_0}^2$ is to be performed.
The Cauchy-Schwarz inequality (\ref{eqn:CS2}) is saturated along the parabolic boundary curve (\ref{eq:constraint2}), while the inequalities in Eq.\ (\ref{eq:ineqcov}) are saturated along the straight line (\ref{eq:constraint4}) connecting $C_+$ and $C_-$ given in Eqs.\ (\ref{eq:AB1}) and (\ref{eq:AB2}).
(b) The relationship (\ref{eqn:Linear}) between $m$ and $n$, when the Cauchy-Schwarz inequality (\ref{eqn:CS2}) is saturated, along the parabolic boundary curve in (a).}
\label{fig:xyplane}
\end{figure}
Another bound on $x$ and $y$ is obtained using the inequality between arithmetic and geometric means, combined with the constraint (\ref{eqn:Ave}), i.e.,
\begin{equation}
\Ave[m]\Ave[n] \leq \left(\frac{\Ave[m]+\Ave[n]}{2}\right)^2 = N^2/4.
\label{eqn:IneqArithGeo}
\end{equation}
Recalling the definition of the covariance and the fact that $mn\ge0$ we find 
\begin{align}
z=\Cov[m,n]
&=\Ave[mn]-\Ave[m]\Ave[n]
\nonumber\\
&\ge -\Ave[m]\Ave[n]\ge -N^2/4.
\label{eq:ineqcov}
\end{align}
Combined with Eq.\ (\ref{eq:var}) it bounds $x$ and $y$ as
\begin{equation}\label{eq:constraint3}
x+y \leq N^2/2+\Delta N^2.
\end{equation}
Its boundary is a line on the $xy$-plane,
\begin{equation}\label{eq:constraint4}
x+y = N^2/2+\Delta N^2,
\end{equation}
intersecting  the $x$ and $y$ axes at $(N^2/2+\Delta N^2,0)$ and $(0,N^2/2+\Delta N^2)$, respectively. 
See Fig.\ \ref{fig:xyplane}(a).

Summarizing, the maximization of $h^2$ in Eq.\ (\ref{eq:h2}) has to be carried out in the domain whose boundary is defined by the parabola (\ref{eq:constraint2}) and the line (\ref{eq:constraint4}) as shown in Fig.\ \ref{fig:xyplane}(a). 
{The corner points $C_\pm$ are given by 
\begin{equation}\label{eq:AB1}
C_\pm=(N^2\sigma_\pm/4,N^2\sigma_\mp/4),
\end{equation}
with
\begin{equation}\label{eq:AB2}
\sigma_\pm=
1+\frac{2\Delta N^2}{N^2}\pm\sqrt{\left(1+\frac{2\Delta N^2}{N^2}\right)^2-1}.
\end{equation}

{It is easy to see that, within this domain, $h^2$ does not have a stationary point.
Indeed, by direct calculation we find
\begin{equation}
 \frac{\partial h^2}{\partial x}=\varepsilon_+(\varepsilon_+-\varepsilon_-),\quad \frac{\partial h^2}{\partial y}=-\varepsilon_-(\varepsilon_+-\varepsilon_-),
\end{equation} 
which cannot vanish simultaneously since we are assuming $\varepsilon_+\neq\varepsilon_-$.
Therefore, the maximum of $h^2$ must be searched on the boundary of the domain.
Let us hence evaluate $h^2$ in Eq.\ (\ref{eq:h2}) along the parabola (\ref{eq:constraint2}), by removing $x+y$ from $h^2$ in Eq.\ (\ref{eq:h2}) using the parabola (\ref{eq:constraint2}), to get
\begin{equation}
h^2_\text{para}
=
\frac{(\varepsilon_+-\varepsilon_-)^2}{\Delta N^2}\xi^2
+(\varepsilon_+^2-\varepsilon_-^2)\xi
+\frac{1}{4}(\varepsilon_++\varepsilon_-)^2\Delta N^2,
\label{eq:parares}
\end{equation} 
where $\xi=(x-y)/2$.
Due to the convexity of Eq.\ (\ref{eq:parares}) as a function of $\xi$, we have that
\begin{equation}\label{eq:maxima}
 \max_P h^2_\text{para}=\max\{h^2_+,h^2_-\}
= h_+^2,
\end{equation}
where $h^2_\pm$ are the values of $h^2$ at the end points $C_\pm$.
If we perform the same analysis for the line defined in Eq.\ (\ref{eq:constraint4}), we find
\begin{equation}
h^2_\text{line}=(\varepsilon_+^2-\varepsilon_-^2)\xi
+\frac{1}{4}(\varepsilon_+-\varepsilon_-)^2N^2
+\frac{1}{2}(\varepsilon_+^2+\varepsilon_-^2)\Delta N^2.
\label{eq:h2line}
\end{equation} 
If $\varepsilon_+^2 =\varepsilon_-^2$, then $h_\text{line}^2$ is constant along the segment (\ref{eq:constraint4}) between $C_+$ and $C_-$.
Otherwise, $h_\text{line}^2$ is maximal at the end point $C_+$, i.e.,
\begin{equation}
\max_P h^2_\text{line} = h^2_+,
\end{equation}
as in Eq.\ (\ref{eq:maxima}).

Finally, by inserting Eqs.\ (\ref{eq:AB1}) and (\ref{eq:AB2}) into Eq.\ (\ref{eq:parares}) or Eq.\ (\ref{eq:h2line}), we see that 
in any case (including the case $\varepsilon_+^2 =\varepsilon_-^2$), we have 
\begin{align}
&\max_P h^2
\nonumber\\
&\ \ =\frac{1}{4}N^2\Biggl(
|\varepsilon_+-\varepsilon_-|\sqrt{1+\left(\frac{\Delta N}{N}\right)^2}
+|\varepsilon_++\varepsilon_-|\frac{\Delta N}{N}
\Biggr)^2,
\label{eqn:h2opt}
\end{align}
which implies, by (\ref{eq:small_c_d_h}) and~(\ref{eqn:FQdef}),
\begin{align}
F_Q^{(\text{max})}(\varphi)
&=\left(
|\varepsilon_+-\varepsilon_-|\sqrt{N^2+\Delta N^2}
+|\varepsilon_++\varepsilon_-|\Delta N
\right)^2  
\nonumber\\
&=\Bigl(
\sqrt{(A_++A_-)^2 + 4|B|^2}\sqrt{N^2+\Delta N^2}
\nonumber\\[-2truemm]
&\qquad\qquad\qquad\qquad\qquad\quad
{}+|A_++A_-|\Delta N
\Bigr)^2.
\label{eq:central}
\end{align}
This formula is the central result of the paper. It shows the dependence of the maximal QFI on the average $N$ and the variance $\Delta N^2$ of the total number of probe particles. 
Moreover, on one hand, if we fix the number of particles at $N$, and thus set $\Delta N=0$, we recover the Heisenberg limit found in Eq.\ (\ref{eqn:QFINOON}).
On the other hand, the fluctuation $\Delta N$ enhances the QFI, and for $\Delta N\gg N$, we get
\begin{equation}
F_Q^{(\text{max})}(\varphi) 
= 4 \varepsilon_+^2 \Delta N^2,
\end{equation}
independent of the average number of particles $N$.

\section{Optimal Input State}\label{sec:OptimalState}
In this  section we are going to determine
the input state $|\psi_0\rangle$ that yields the maximum value of the QFI (\ref{eq:central}) obtained in the previous section.

\subsection{For $\bm{\varepsilon_++\varepsilon_-\neq0}$: Quasi-NOON State}\label{sec:quasinoon}
We recall that, when the maximum of $h^2$ in Eq.\ (\ref{eqn:h2opt}) is achieved, the inequalities in Eq.\ (\ref{eq:ineqcov}) are saturated.
This implies [recall the condition for the equality between the arithmetic and geometric means in (\ref{eqn:IneqArithGeo})]
\begin{equation}
\Ave[mn]=0,\qquad
\Ave[m]=\Ave[n]=N/2,\label{eqn:Emn}
\end{equation}
and hence, $\Cov[m,n]$ and the constraint (\ref{eq:var}) are reduced to
\begin{equation}
\Cov[m,n]
=-N^2/4,\ \ %
\Var[m]+\Var[n] = N^2/2+\Delta N^2.\label{eqn:CovN2}
\end{equation}
When $\varepsilon_++\varepsilon_-\neq0$ 
the Cauchy-Schwarz inequality (\ref{eqn:CS2}), i.e., $|{\Cov[m,n]}|\le\sqrt{\Var[m]\Var[n]}$, is also saturated.
This implies (recall the condition for the saturation of the Cauchy-Schwarz inequality) that the random variables $m$ and $n$ are linearly dependent,
\begin{equation}
n=\alpha m+\beta\qquad(\alpha<0),
\label{eqn:Linear}
\end{equation}
with {$\alpha$ and $\beta$ real numbers.}
Due to the negative covariance in Eq.\ (\ref{eqn:CovN2}), the coefficient $\alpha$ is negative, and the ranges of $m$ and $n$ are limited:
see Fig.\ \ref{fig:xyplane}(b).
In addition, we know that the maximum of $h^2$ is reached at 
the end point $C_+$ in Eq.\ (\ref{eq:AB1}) [since $\varepsilon_+^2>\varepsilon_-^2$: see Eq.\ (\ref{eq:maxima})], i.e., when
\begin{equation}
(\Var[m],\Var[n])=
(N^2\sigma_+/4,N^2\sigma_-/4),
\label{eqn:VarVar}
\end{equation}
where $\sigma_\pm$ are defined in Eq.\ (\ref{eq:AB2}).
From these conditions, $\alpha$ and $\beta$ are fixed.
By inserting Eq.\ (\ref{eqn:Linear}) into Eqs.\ (\ref{eqn:Emn}) and (\ref{eqn:CovN2}), we have
\begin{equation}
N\alpha+2\beta=N,\qquad
\alpha\Var[m]=-N^2/4,
\end{equation}
and, by taking Eq.\ (\ref{eqn:VarVar}) into account, we get
\begin{equation}
\alpha=-\sigma_-,\qquad
\beta=N(1+\sigma_-)/2.
\end{equation}
It derives that the ranges of the random variables $m$ and $n$ are limited by
\begin{equation}
0\le m\le N(1+\sigma_+)/2,\quad
0\le n\le N(1+\sigma_-)/2.
\end{equation}
We are now ready to characterize the structure of the optimal input state that maximizes $h^2$ for the case $\varepsilon_++\varepsilon_-\ne 0$.
Due to the linear dependence (\ref{eqn:Linear}), the random variables $m$ and $n$ are perfectly correlated, and we have only to look 
for a probability distribution $P_m$ for the variable $m$.
The first condition in Eq.\ (\ref{eqn:Emn}) implies $mn=0$, and, therefore, either $m$ or $n$ should vanish.
When $n=0$ we have 
$m=N(1+\sigma_+)/2$.
Thus, the solution is
\begin{equation}
P_m=\lambda \delta_{m,N(1+\sigma_+)/2}+(1-\lambda)\delta_{m,0}
\quad(0\le\lambda\le1).
\label{eqn:Pm}
\end{equation}
This probability distribution should satisfy the conditions in Eqs.\ (\ref{eqn:Emn}) and (\ref{eqn:VarVar}), i.e.,
\begin{equation}
\begin{cases}
\medskip
\displaystyle
\Ave[m]=\lambda N(1+\sigma_+)/2=N/2,\\
\displaystyle
\Var[m]=\lambda (1-\lambda )N^2(1+\sigma_+)^2/4=N^2\sigma_+/4,
\end{cases}
\end{equation}
which are compatible with each other with the choice
\begin{equation}
\lambda=\frac{1}{1+\sigma_+}.
\end{equation}
Therefore, the optimal input state takes the form
\begin{align}
\ket{\psi_0}
={}&\sqrt{\frac{1}{1+\sigma_+}}\ket{N(1+\sigma_+)/2,0}
\nonumber\\
&{}+{\sqrt{\frac{1}{1+\sigma_-}}}e^{i\phi}\ket{0,N(1+\sigma_-)/2},
\label{eqn:NOON-like}
\end{align}
where the phase $\phi$ {is} arbitrary. This state is a deformation of  a NOON state  with different weights (\emph{quasi-NOON state}). 
In the limit of vanishing fluctuation $\Delta N/N\to0$, we have $\sigma_\pm\to1$, and the state (\ref{eqn:NOON-like}) reduces to the NOON state (\ref{eqn:NOON}).

Notice that the variables $m$ and $n$ are integers  while they should take the values $m=N(1+\sigma_+)/2$ and $n=N(1+\sigma_-)/2$ in the state (\ref{eqn:NOON-like}), which are in general nonintegers.
Precisely speaking, the true optimal state is different from the one given in Eq.\ (\ref{eqn:NOON-like}) in such a generic case, and the QFI given in Eq.\ (\ref{eq:central}) is not reachable.
The error, however, is negligibly small, of order $1/N$, when $N$ is large. (Moreover, this problem can be removed with a proper choice of the ratio $\Delta N/N$.)

Note also that the optimal state (\ref{eqn:NOON-like}) is a quasi-NOON state defined in the basis diagonalizing the generator $\hat{H}_\varphi$ as Eq.\ (\ref{eqn:Hcd}), i.e., in the normal modes $\hat{c}_\pm$, not  in the physical modes $\hat{a}_\pm$.
Since the unitary transformation relating the physical modes $\hat{a}_\pm$ to the normal modes $\hat{c}_\pm$ is a unitary transformation preserving the total number of particles as Eq.\ (\ref{enne}), such a transformation can in practice be realized by a linear circuit composed of beam splitters and phase shifters.
Therefore, the optimal state $\ket{\psi_0}$ given in Eq.\ (\ref{eqn:NOON-like}) to be injected into the target circuit can be generated from the quasi-NOON state prepared in the physical modes $\hat{a}_\pm$ by sending it through an appropriate linear circuit realizing the diagonalization of the generator $\hat{H}_\varphi$.
Recall the formula for the QFI in Eq.\ (\ref{eqn:FQdef}) with Eq.\ (\ref{variance0}), and consider the diagonalization of the generator $\hat{H}_\varphi$ in that expression, to understand that the change of basis for the generator $\hat{H}_\varphi$ induces the unitary transformation on the input state $\ket{\psi_0}$.

It is worth mentioning that in order to construct the optimal state one needs some knowledge on the parameter $\varphi$, which can be approximately learned via the following strategy: an initial small fraction of probes is used to acquire preliminary information on the value of $\varphi$ in order to prepare the optimal input state for the subsequent probes.
This will add an extra cost to the procedure that, however, will be asymptotically negligible for large enough $N$.

\subsection{For $\bm{\varepsilon_++\varepsilon_-=0}$: Two-Mode Schr\"odinger-Cat State with Vacuum}\label{sec:schcat}
In the case of an antisymmetric scatterer with $\varepsilon_++\varepsilon_-=0$ [e.g., the Mach-Zehnder interferometer in Fig.\ \ref{fig:BlackBox}(b), for which we have $\varepsilon_\pm=\pm1/2$], the maximum of $h^2$ in Eq.\ (\ref{eqn:h2opt}) is reached anywhere on the segment with end points $C_+$ and $C_-$ in Fig.\ \ref{fig:xyplane}(a).
In this case the inequalities in Eq.\ (\ref{eq:ineqcov}) are saturated, and the conditions (\ref{eqn:Emn}) and (\ref{eqn:CovN2}) are satisfied, but the Cauchy-Schwarz inequality (\ref{eqn:CS2}) is not necessarily saturated.
We do not have the perfect correlation (\ref{eqn:Linear}) between $m$ and $n$ in general.
Still, we have a strong constraint, i.e., the first one in Eq.\ (\ref{eqn:Emn}), which implies that either $m$ or $n$ should vanish.
Therefore, the relevant probability distribution that maximizes $h^2$ for $\varepsilon_++\varepsilon_-=0$ is given in the form
\begin{equation}
P_{m,n}
=\mu\delta_{n,0}p_m
+(1-\mu)\delta_{m,0}\tilde{p}_n
\quad(0\le\mu\le1),
\label{eqn:Pmn}
\end{equation}
where $p_m$ and $\tilde{p}_n$ are normalized probability distributions.
This probability $P_{m,n}$ should satisfy the conditions in Eqs.\ (\ref{eqn:Emn}) and (\ref{eqn:CovN2}), i.e.,
\begin{equation}
\begin{cases}
\medskip
\displaystyle
\Ave[m]=\Ave[n]=N/2,
\\
\Ave[m^2] + \Ave[n^2] = N^2+\Delta N^2,
\end{cases}
\label{eqn:Emn2}
\end{equation}
namely,
\begin{equation}
\begin{cases}
\medskip
\displaystyle
\mu\sum_mmp_m
=(1-\mu)\sum_nn\tilde{p}_n
=N/2,
\\
\displaystyle
\mu\sum_mm^2p_m+(1-\mu)\sum_nn^2\tilde{p}_n=N^2+\Delta N^2.
\end{cases}
\label{eqn:pmmpnn}
\end{equation}
There are infinitely many triples $(p_m,\tilde{p}_n,\mu)$ satisfying these conditions, among which we have the above two explicit examples (\ref{eqn:Pm}), i.e., $p_m=\delta_{m,N(1+\sigma_+)/2}$ and $\tilde{p}_n=\delta_{n,N(1+\sigma_-)/2}$  with $\mu=1/\sqrt{1+\sigma_+}$.
That is, the quasi-NOON states in Eq.\ (\ref{eqn:NOON-like}) give the optimal QFI also in this case.
Another nontrivial example can be constructed with Poissonian distributions
\begin{gather}
p_m=\frac{1}{m!}\left(\frac{N}{2\mu_+}\right)^me^{-N/2\mu_+},
\label{eqn:Poissonian}\\
\tilde{p}_n=\frac{1}{n!}\left(\frac{N}{2\mu_-}\right)^ne^{-N/2\mu_-},\\
\mu_\pm=\frac{1}{2}\left(
1\pm\sqrt{\frac{\Delta N^2-N}{N^2+\Delta N^2-N}}
\right), \quad \mu=\mu_+,
\label{eqn:PoissonianMu}
\end{gather}
as long as $\Delta N^2\ge N$.
The set of probability distributions $\{P_{m,n}\}$ satisfying the above conditions in Eq.\ (\ref{eqn:pmmpnn}) forms a convex set: any convex combinations of valid probability distributions $P_{m,n}$ and $P'_{m,n}$ are also valid probability distributions, since the conditions in Eq.\ (\ref{eqn:Emn2}) are linear in $P_{m,n}$.

Each probability distribution (\ref{eqn:Pmn}) can be arranged as
\begin{equation}
P_{m,n}
=\delta_{m,0}\delta_{n,0}[\mu p_0+(1-\mu)\tilde{p}_0]
+\mu\delta_{n,0}q_m
+(1-\mu)\delta_{m,0}\tilde{q}_n,
\end{equation}
where $q_m=p_m-p_0\delta_{m,0}$ and $\tilde{q}_n=\tilde{p}_n-\tilde{p}_0\delta_{n,0}$, and the corresponding pure state reads
\begin{align}
\ket{\psi_0}
={}&\sqrt{\mu p_0+(1-\mu)\tilde{p}_0}\ket{0,0}
\nonumber\\[2truemm]
&{}
+\sum_{m>0}\sqrt{\mu p_m}\,e^{i\phi_m}\ket{m,0}
\nonumber\\
&{}+\sum_{n>0}\sqrt{(1-\mu)\tilde{p}_n}\,e^{i\tilde{\phi}_n}\ket{0,n}, \label{stato1} 
\end{align}
with arbitrary phases $\phi_m$ and $\tilde{\phi}_n$. As a special instance of such optimal  states, it is worth considering the case with the Poissonian distributions (\ref{eqn:Poissonian})--(\ref{eqn:PoissonianMu}) with the phases $\phi_m$ and 
$\tilde{\phi}_n$ having linear dependences on their indices, i.e., 
\begin{equation} 
\phi_m = m \Delta \phi + \phi_0 , \qquad \tilde{\phi}_n = n \Delta \tilde{\phi} + \tilde{\phi}_0,
\end{equation} 
where $\Delta \phi$, $\phi_0$,  $\Delta \tilde{\phi}$, and $\tilde{\phi}_0$ are arbitrary constants. 
Under this assumption, Eq.\ (\ref{stato1}) can be expressed as a coherent superposition of three two-mode coherent states,
\begin{equation}
\ket{\psi_0}
= c_0 
\ket{0,0} + \sqrt{\mu} \, e^{i \phi_0} \ket{\alpha,0} + \sqrt{1-\mu}\,  e^{i \tilde{\phi}_0} \ket{0,\tilde{\alpha}},
\label{stato111} 
\end{equation}
where the probability amplitude associated with the vacuum term is given by 
\begin{align} 
c_0 ={}& \sqrt{\mu p_0+(1-\mu)\tilde{p}_0}
\nonumber\\
&{} - \sqrt{\mu e^{-N/2\mu_+} }\,e^{i \phi_0} - \sqrt{(1- \mu) e^{-N/2\mu_-} }\,e^{i \tilde{\phi}_0} , 
\end{align}
while the amplitudes $\alpha$ and $\tilde{\alpha}$ of the coherent states $\ket{\alpha,0}$ and $\ket{0,\tilde{\alpha}}$ are 
\begin{equation} 
\alpha= \sqrt{\frac{N}{2\mu_+}}e^{i \Delta \phi}, \qquad \tilde{\alpha}=  \sqrt{\frac{N}{2\mu_-}} e^{i \Delta \tilde{\phi}}.
\end{equation}
It is worth observing that the vector (\ref{stato111}) is a two-mode Schr\"odinger-cat state with three components, one being the vacuum and the other two having a mean photon number $|\alpha|^2$ in mode ``$+$'' and a mean photon number $|\tilde{\alpha}|^2$ in mode ``$-$'' both of which scale linearly with $N$. 
Note again that the ``modes'' here are not the physical modes $\hat{a}_\pm$ but the normal modes $\hat{c}_\pm$.

\section{Special Cases}\label{sec:special}
In this section we discuss explicitly three paradigmatic  examples of scattering processes, relevant to our previous discussion.

The first setting  is an example of an asymmetric configuration: see  
 Fig.\ \ref{fig:SpecialCases}(i).  
 It is equivalent to the Mach-Zehnder interferometer in Fig.\ \ref{fig:BlackBox}(b) (but with $\varphi/2$ replaced by $\varphi$), apart from the first and the last beam splitters, which are both irrelevant to the QFI and are used to change the input state and the measurement procedure.  
In this scheme the probes in the two input modes experience opposite phase shifts. Accordingly, the generator $\hat{H}_\varphi$ (i.e., $\hat{a}_\pm= \hat{c}_\pm$) is already in the normal form (\ref{eqn:Hcd}) with $(\varepsilon_+,\varepsilon_-)=(1,-1)$ yielding 
\begin{align}
&F_Q^\text{(max)}(\varphi) = 4 (N^2+\Delta N^2) 
\nonumber\\
&\qquad \Longrightarrow \quad   \delta \varphi_\text{min} = \frac{1}{2\sqrt{\nu (N^2+\Delta N^2) }}, 
  \label{PHIMIN1}
\end{align} 
obtained by either quasi-NOON input states (\ref{eqn:NOON-like}) or (when $\Delta N^2 \geq N$)  the vectors (\ref{stato1}).

\begin{figure}[b]
\begin{tabular}{l@{\ \ }l@{\ \ }l}
(i)&(ii)&(iii)\\
\makebox(76,39)[t]{\includegraphics[width=0.15\textwidth]{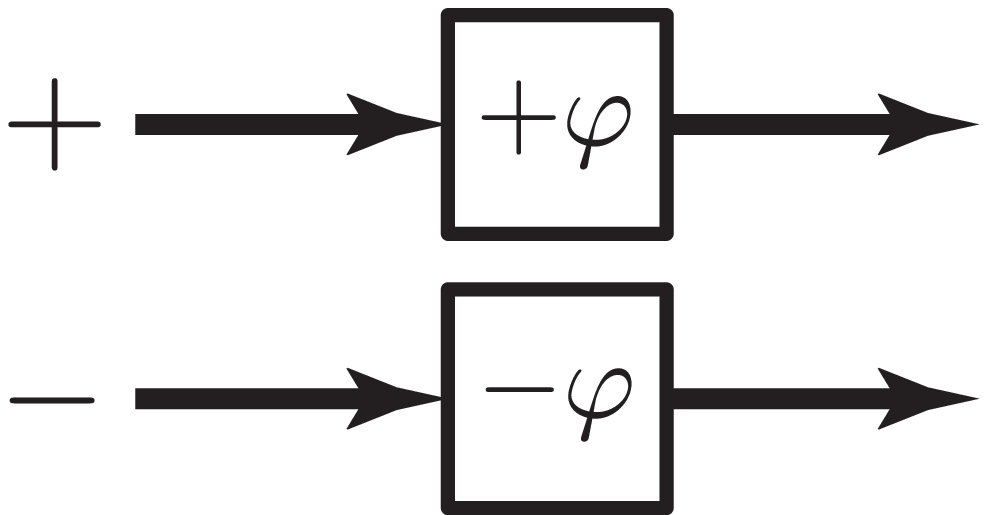}}&
\makebox(76,39)[t]{\includegraphics[width=0.15\textwidth]{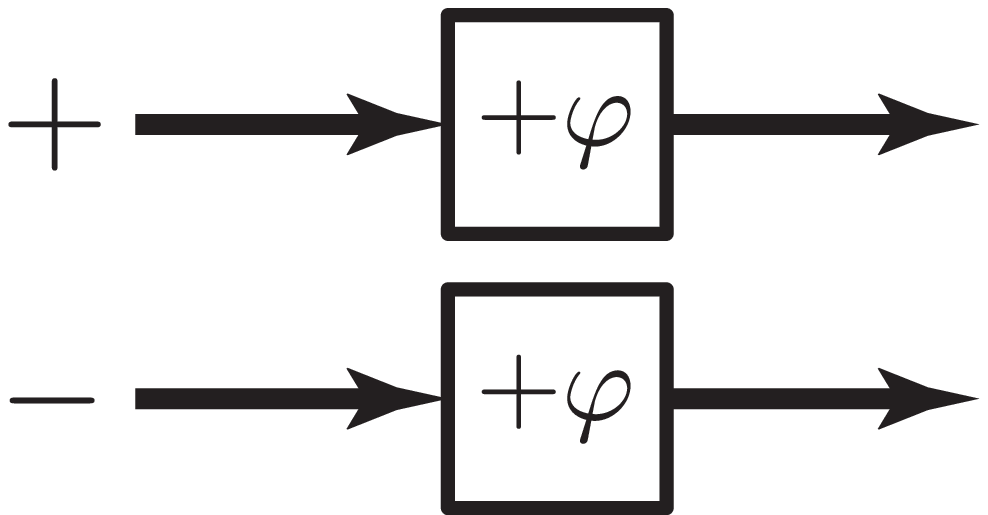}}&
\makebox(76,39)[t]{\includegraphics[width=0.15\textwidth]{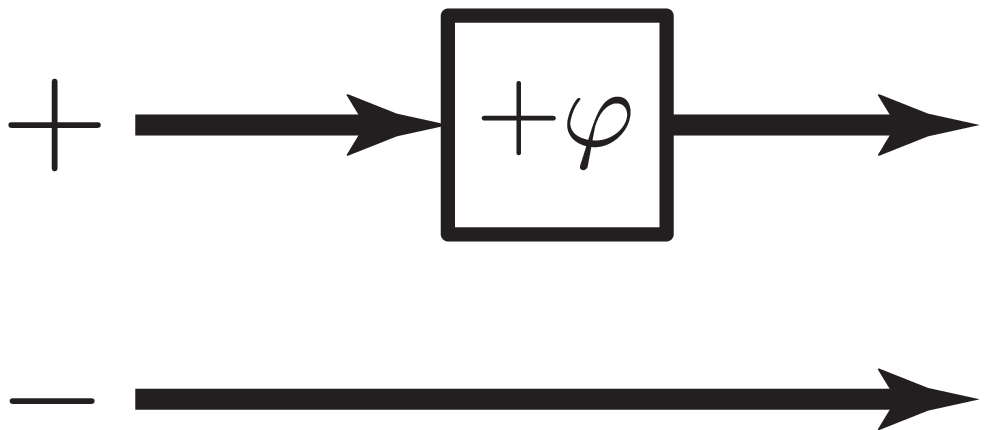}}
\end{tabular}
\caption{Three paradigmatic settings for the estimation of a phase shift $\varphi$: (i) antisymmetric configuration; (ii) symmetric configuration; (iii) unbalanced configuration.
In all cases the input modes  $\hat{a}_\pm$ coincide  with the normal modes $\hat{c}_\pm$ of the system.  }
\label{fig:SpecialCases}
\end{figure}

The second setting  also involves two phase shifters, but in the parallel configuration: see  
 Fig.\ \ref{fig:SpecialCases}(ii). This is an example of a symmetric setup where the generator
 $\hat{H}_\varphi$ is again  in the normal form (\ref{eqn:Hcd}) with now  $(\varepsilon_+,\varepsilon_-)=(1,1)$. Accordingly, we get 
 \begin{equation}  \label{PHIMIN2}
F_Q^\text{(max)}(\varphi) = 4\,\Delta N^2  \quad \Longrightarrow \quad   \delta \varphi_\text{min} = \frac{1}{2\sqrt{\nu}  \Delta N}, 
  \end{equation} 
the optimal state being any state fulfilling the particle variance constraint.

Finally, the last setting  is an example of an unbalanced scatterer, where the probe in one arm does not encounter a phase shifter and just acts as a reference: see  
 Fig.\ \ref{fig:SpecialCases}(iii). Also in this case
$\hat{H}_\varphi$ is  in the normal form (\ref{eqn:Hcd}) with  $(\varepsilon_+,\varepsilon_-)=(1,0)$ yielding 
\begin{align}
&F_Q^\text{(max)}(\varphi) = (
\sqrt{N^2+\Delta N^2}
+\Delta N
)^2
\nonumber\\
&\qquad
\Longrightarrow \quad
 \delta \varphi_\text{min} = \frac{1}{\sqrt{\nu}  (
\sqrt{N^2+\Delta N^2}
+\Delta N
) }, 
  \label{PHIMIN3}
\end{align} 
  the optimal probes  being the quasi-NOON input states in Eq.\ (\ref{eqn:NOON-like}).

It is interesting to observe that, for fixed values of the constraints $N$ and $\Delta N$,  the antisymmetric configuration (i) yields the best estimation uncertainty $\delta \varphi_\text{min}$, the  unbalanced configuration (iii) gives the second best uncertainty, while  the parallel configuration (ii) is the worst of the three.

\section{Optimal Gaussian Input States}\label{sec:OptGauss}
In the previous section, we have identified the optimal input states that allow us to achieve the optimal QFI in Eq.\ (\ref{eq:central}), for the generic two-mode linear circuits, with a given average and variance of the total number of probe bosons.
In particular, the quasi-NOON state (\ref{eqn:NOON-like}) is the optimal choice if $\varepsilon_+\neq\varepsilon_-$.
If $\varepsilon_+=-\varepsilon_-$ and $\Delta N^2\ge N$, the Schr\"odinger-cat state with vacuum in Eq.\ (\ref{stato111}) is one among infinitely many optimal states.
However, generating such exotic states  might be very challenging in practice, even in quantum optical implementations \cite{ref:NOON2002-KokLeeDowling,ref:NOON2003-PrydeWhite,ref:NOON2004-Walther,ref:NOON2004-Mitchell,ref:MetrologyExpTakeuchi,ref:NOON2007-PrydeOBrienWhite,ref:Cable-PRL,ref:NOON2010-Afek,ref:NOON2012-Israel}.
On the contrary, Gaussian states, including entangled states such as two-mode squeezed states, appear to be much easier to produce, and to some extent are readily available in the laboratory \cite{ref:ContVarQI,ref:GuassianQI}:  identifying the optimal Gaussian input states which provide the best performance in our setting
hence appears to be an interesting question.
 
Quantum metrology using Gaussian states of light, in particular with two-mode interferometric setups, has also been eagerly studied in the literature \cite{DOWLING,ref:DowlingParity,CAVES,ref:caves1,ref:caves2,PEZZE1}.
We notice here that since the quasi-NOON states (\ref{eqn:NOON-like}) and the Schr\"odinger-cat states (\ref{stato111}) are not Gaussian states, the optimal QFI in Eq.\ (\ref{eq:central}) cannot be reached by a Gaussian state in general.
How much is the gap between the ultimate QFI in Eq.\ (\ref{eq:central}) and the one attainable by the best Gaussian state?
What is the structure of the best Gaussian state?
These are the questions we address in this section. For this purpose, however,  we observe that in general 
it is not easy to variate the parameters characterizing Gaussian states so that the constraints on the average $N$ and the variance $\Delta N^2$ of the number of probe particles are always satisfied simultaneously.
On the other hand, due to the Gaussianity of the probe state, fixing the average at $N$ is enough to get a finite optimal QFI\@.
We therefore search for the optimal input state maximizing the QFI for the estimation of a parameter $\varphi$ of the generic two-mode linear circuit in Fig.\ \ref{fig:BlackBox}(a), among the Gaussian states fulfilling only the constraint on the average number $N$ of probe particles.

\subsection{QFI with a Gaussian Input State}
Each Gaussian state is characterized by the covariance {matrix $\Gamma$ defined as
\begin{equation}
\Gamma_{ij}
=\frac{1}{2}\langle\hat{z}_i\hat{z}_j+\hat{z}_j\hat{z}_i\rangle
-\langle\hat{z}_i\rangle\langle\hat{z}_j\rangle\quad
(i,j=1,\ldots,4),
\end{equation} 
with $\langle\,\cdots\,\rangle$ denoting the expectation value in 
the Gaussian state \cite{ref:ContVarQI,ref:GuassianQI}, where we have used the quadrature operators
\begin{equation}
\hat{\bm{z}}
=\begin{pmatrix}
\hat{x}_+\\
\hat{x}_-\\
\hat{y}_+\\
\hat{y}_-
\end{pmatrix},\quad
\hat{x}_\pm=\frac{\hat{a}_\pm+\hat{a}_\pm^\dag}{\sqrt{2}},\ \ %
\hat{y}_\pm=\frac{\hat{a}_\pm-\hat{a}_\pm^\dag}{\sqrt{2}\,i},
\label{eqn:quadrature}
\end{equation}
and the displacement $\bm{d}$, whose components are defined as
\begin{equation}
d_i=\langle\hat{z}_i\rangle\qquad
(i=1,\ldots,4).
\end{equation}}

We focus on pure input states, as in the optimization problem studied in the previous sections.
In this case all the  eigenvalues of $\Gamma$ (obtained using Williamson's theorem \cite{ref:williamson}) are equal to $1/2$, and $\Gamma$ can always be decomposed as \cite{ref:ContVarQI,ref:GuassianQI}
\begin{equation}
\Gamma
=\frac{1}{2}RQ^2R^T,
\label{eqn:GammaDecomp}
\end{equation}
where $Q=\diag(e^{r_+},e^{r_-},e^{-r_+},e^{-r_-})$ is the (diagonal) squeezing matrix and $R$ is a symplectic and orthogonal (thus, unitary) matrix. Due to its symplectic structure, $R$ can be written as
\begin{equation}
R=W^\dag
\left(
\begin{array}{c|c}
\,U\,&0\\
\hline
0&U^*
\end{array}
\right)
W,\quad
W=\frac{1}{\sqrt{2}}
\left(
\begin{array}{c|c}
\,\,\mathbb{I}_{2}\,\,&i\mathbb{I}_{2}\\
\hline
\mathbb{I}_{2}&-i\mathbb{I}_{2}
\end{array}
\right),
\label{eqn:R}
\end{equation}
where $U$ is a $2\times2$ unitary matrix ($U^*$ being its conjugate), $W$ is a $4\times4$ unitary matrix and $\mathbb{I}_{2}$ is the $2\times2$ identity matrix.

When injecting a pure Gaussian state characterized by a covariance matrix $\Gamma$ and a displacement $\bm{d}$ into the two-mode linear circuit in Fig.\ \ref{fig:BlackBox}(a)
the associated linear
  transformation (\ref{eqn:Smatrix}) maps it  into a new Guassian state with a covariance matrix and displacement given by 
  \begin{equation}
\Gamma_\varphi=R_\varphi\Gamma R_\varphi^T,\qquad
\bm{d}_\varphi
=R_\varphi\bm{d},
\label{eqn:GammaDRotAgain2}
\end{equation}
where $R_\varphi$ is the unitary ($\varphi$-dependent) matrix defined by 
 \begin{equation}
R_\varphi=W^\dag
\left(
\begin{array}{c|c}
U_\varphi&0\\
\hline
0&U^*_\varphi
\end{array}
\right)
W,
\label{eqn:GammaDRotAgain}
\end{equation}
where $U^*_\varphi$ is the $2\times2$ matrix in Eq.\ (\ref{eqn:Smatrix}) and $W$ is defined as in Eq.\ (\ref{eqn:R}).

The QFI for the pure Gaussian state with a covariance matrix $\Gamma_\varphi$ and a displacement $\bm{d}_\varphi$ can then be computed along the lines of  Ref.\ \cite{ref:Monras-QFIGauss}, yielding 
\begin{align}
F_G(\varphi)
&=\frac{1}{4}\Tr\!\left\{
\left(
\Gamma_\varphi^{-1}
\frac{\partial\Gamma_\varphi}{\partial\varphi}
\right)^2
\right\}
+
\frac{\partial\bm{d}_\varphi^T}{\partial\varphi}
\Gamma_\varphi^{-1}
\frac{\partial\bm{d}_\varphi}{\partial\varphi} 
\nonumber \\
&=\frac{1}{2}\Tr\{
H_\varphi\Gamma^{{-1}}H_\varphi\Gamma
-H_\varphi^2
\}
+\bm{d}^TH_\varphi\Gamma^{{-1}}H_\varphi\bm{d}.
\label{eqn:QFIGaussPureLinear}
\end{align} 
In this expression  the $4\times 4$ matrix $H_\varphi$  is the generator of $R_\varphi$, which,  in a properly chosen basis, is given by 
\begin{equation}
H_\varphi=iR_\varphi^T\frac{\partial R_\varphi}{\partial\varphi}
= W^\dag
\left(\begin{array}{c|c}
\,\,\varepsilon\,\,&0\\
\hline
0&-\varepsilon
\end{array}\right)
W 
,\quad
\varepsilon
=
\begin{pmatrix}
\medskip
\varepsilon_+&0\\
0&\varepsilon_-
\end{pmatrix},
\end{equation}
with $\varepsilon_\pm$ as in Eq.\ (\ref{eqn:Epsilon}): the dependence of $F_G(\varphi)$  upon the parameter $\varphi$ is once more entirely encoded in these 
functions. 
We find it convenient to rewrite Eq.\ (\ref{eqn:QFIGaussPureLinear}) as
\begin{equation}
F_G(\varphi)
=F_G^{(1)}(\varphi)
+F_G^{(2)}(\varphi)
\label{eqn:QFIG}
\end{equation}
with
\begin{widetext}
\begin{equation}
\begin{cases}
\medskip
\displaystyle
F_G^{(1)}(\varphi)
=
\Tr\{
U^\dag\varepsilon U\sinh2r\,(U^\dag\varepsilon U)^*\sinh2r
+(U^\dag\varepsilon U\cosh2r)^2
-\varepsilon^2
\},\\
\displaystyle
F_G^{(2)}(\varphi)
=2
\left(\begin{array}{c|c}
\bm{\alpha}^\dag\varepsilon U&
\bm{\alpha}^T\varepsilon U^*
\end{array}\right)
\left(\begin{array}{c|c}
\cosh2r&
\sinh2r\\
\hline
\sinh2r&
\cosh2r
\end{array}\right)
\left(\begin{array}{c}
U^\dag\varepsilon\bm{\alpha}\\
\hline
U^T\varepsilon\bm{\alpha}^*
\end{array}\right),
\end{cases}
\label{eqn:QFIM}
\end{equation}
\end{widetext}
where we have defined the  $2\times2$ diagonal matrices $\cosh2r=\diag(\cosh2r_+,\cosh2r_-)$, $\sinh2r=\diag(\sinh2r_+,\sinh2r_-)$ and the vector
\begin{equation}
\bm{\alpha}
=\begin{pmatrix}
\medskip
\alpha_+\\
\alpha_-
\end{pmatrix}
=\begin{pmatrix}
\medskip
\langle\hat{a}_+\rangle\\
\langle\hat{a}_-\rangle
\end{pmatrix}
=
\frac{1}{\sqrt{2}}
\begin{pmatrix}
\medskip
\langle\hat{x}_+\rangle+i\langle\hat{y}_+\rangle\\
\langle\hat{x}_-\rangle+i\langle\hat{y}_-\rangle\\
\end{pmatrix}.
\end{equation}
The first contribution $F_G^{(1)}(\varphi)$ to the QFI can be explicitly evaluated parametrizing the $2\times2$ unitary matrix $U$ in Eq.\ (\ref{eqn:R}) as
\begin{equation}
U=e^{-\frac{i}{2}\eta}
\begin{pmatrix}
\medskip
e^{-\frac{i}{2}(\chi+\phi)}\cos\frac{\theta}{2}
&
-e^{\frac{i}{2}(\chi-\phi)}\sin\frac{\theta}{2}
\\
e^{-\frac{i}{2}(\chi-\phi)}\sin\frac{\theta}{2}
&
e^{\frac{i}{2}(\chi+\phi)}\cos\frac{\theta}{2}
\end{pmatrix}.
\label{eq:gaussunitary}
\end{equation}
A direct calculation hence shows that
\begin{align}
F_G^{(1)}(\varphi)
={}&
2
\left(
\bar{\varepsilon}
+\frac{1}{2}\delta\varepsilon\cos\theta
\right)^2
\sinh^22r_+
\nonumber\\
&{}+2
\left(
\bar{\varepsilon}
-\frac{1}{2}\delta\varepsilon\cos\theta
\right)^2
\sinh^22r_-
\nonumber\\[1.5truemm]
&{}+(\delta\varepsilon)^2\sin^2\theta
\,[
\sinh^2(r_++r_-)
\nonumber\\
&\qquad\qquad\qquad\ %
{}-
\sin^2\chi
\sinh2r_+\sinh2r_-
],
\label{eqn:FQ1TwoMode}
\end{align}
where
\begin{equation}
\bar{\varepsilon}=\frac{\varepsilon_++\varepsilon_-}{2},\qquad
\delta\varepsilon=\varepsilon_+-\varepsilon_-. 
\end{equation}

\subsection{Optimization of the Pure Gaussian Input State}\label{sec:gaussstate}
We optimize the QFI in Eq.\ (\ref{eqn:QFIG}), {given} Eqs.\ (\ref{eqn:QFIM}) and (\ref{eqn:FQ1TwoMode}), under the constraint on the average of the number of probe particles $\hat{N}$ defined in Eq.\ (\ref{enne}). In the case of a pure Gaussian state, it can be specialized as
\begin{equation}
\langle\hat{N}\rangle
=\sinh^2r_++\sinh^2r_-+\|\bm{\alpha}\|^2
=N
\label{eqn:ConstraintN}
\end{equation}
with
\begin{equation}
\|\bm{\alpha}\|^2
=\bm{\alpha}^\dag\bm{\alpha}
=|\alpha_+|^2+|\alpha_-|^2.
\end{equation}
Without loss of generality, we take
\begin{equation}
r_+\ge r_-\ge0
\label{eqn:r1r2}
\end{equation}
[if we want other configurations we can just rotate the squeezing matrix $Q$ by $R$ in $\Gamma$ in Eq.\ (\ref{eqn:GammaDecomp})]. Since only the norm {$\|\bm{\alpha}\|^2$} is relevant to the constraint (\ref{eqn:ConstraintN}), we can freely tune the ``direction'' of $\bm{\alpha}$.
Let us first optimize the second contribution $F_G^{(2)}(\varphi)$ to the QFI in Eq.\ (\ref{eqn:QFIM}) by making use of this freedom. The upper bound can be evaluated  using the largest eigenvalue of the matrix in the definition of $F_G^{(2)}(\varphi)$ in Eq.\ (\ref{eqn:QFIM}). We obtain
\begin{equation}
F_G^{(2)}(\varphi)
\le4\|\varepsilon\bm{\alpha}\|^2e^{2r_+}.
\label{eqn:FQ2TwoMode}
\end{equation}
This bound can be made more helpful taking into account that
\begin{align}
\|\varepsilon\bm{\alpha}\|^2
={}&\varepsilon_+^2|\alpha_+|^2
+\varepsilon_-^2|\alpha_-|^2
\nonumber\\
={}&
(\varepsilon_+^2+\varepsilon_-^2)\frac{|\alpha_+|^2+|\alpha_-|^2}{2}
\nonumber\\
&{}+(\varepsilon_+^2-\varepsilon_-^2)\frac{|\alpha_+|^2-|\alpha_-|^2}{2}
\le
\varepsilon_+^2 \|\bm{\alpha}\|^2.
\end{align}
Finally we find
\begin{equation}
F_G^{(2)}(\varphi)
\le4 \varepsilon_+^2 \|\bm{\alpha}\|^2e^{2r_+}.
\label{eqn:FQ2Opt}
\end{equation}
This bound can be saturated by tuning the parameters of the unitary matrix $U$ in Eq.\ (\ref{eq:gaussunitary}) characterizing the input Gaussian state so that
\begin{equation}
\begin{cases}
\medskip
\displaystyle
\bm{\alpha}
=
\pm 
\|\bm{\alpha}\|
e^{-\frac{i}{2}(\eta+\chi+\phi)}
\begin{pmatrix}
\smallskip
1\\
0
\end{pmatrix}
\ \text{and}\ %
\theta=0
&(\varepsilon_+^2>\varepsilon_-^2),\\
\displaystyle
\bm{\alpha}
=\pm
\|\bm{\alpha}\|
e^{-\frac{i}{2}(\eta+\chi)}
\begin{pmatrix}
\smallskip
e^{-\frac{i}{2}\phi}\cos\frac{\theta}{2}
\\
e^{\frac{i}{2}\phi}\sin\frac{\theta}{2}
\end{pmatrix}
&(\varepsilon_+^2=\varepsilon_-^2).
\end{cases}
\label{eqn:Opt2}
\end{equation}
We turn now to the maximization of $F_G^{(1)}(\varphi)$. We notice that the parameters $\theta$ and $\chi$ in Eq.\ (\ref{eqn:FQ1TwoMode}) are also irrelevant to the constraint (\ref{eqn:ConstraintN}).
We next tune them to optimize the first contribution $F_G^{(1)}(\varphi)$ to the QFI in Eq.\ (\ref{eqn:FQ1TwoMode}).
Setting $\chi=0$, $F_G^{(1)}(\varphi)$ is a convex parabolic function of $\cos\theta$ and reaches its maximum at either end of the range of $\cos\theta$, i.e., at $\cos\theta=\pm 1$.
The optimal choice is
\begin{equation}
\theta=0 ,
\label{eqn:ThetaTune}
\end{equation}
and a direct calculation shows that
\begin{equation}
\max_{\theta}
F_G^{(1)}(\varphi)
=2(\varepsilon_+^2
\sinh^22r_+
+
\varepsilon_-^2
\sinh^22r_-),
\label{eqn:FQ1OptAngles}
\end{equation}
irrespective of $\chi$.
The optimal value of $\theta$ for $F_G^{(1)}(\varphi)$ in Eq.\ (\ref{eqn:ThetaTune}) are compatible with those obtained for the maximization of $F_G^{(2)}(\varphi)$ in Eq.\ (\ref{eqn:Opt2}). 

Finally, optimizing also with respect to $\bm{n}=\bm{\alpha}/\|\bm{\alpha}\|$, the direction of $\bm{\alpha}$, we get
\begin{align}
&\max_{\theta,\,\bm{n}} F_G(\varphi)
\nonumber\\
&\ %
=2
[
\varepsilon_+^2
(\sinh^22r_++2\|\bm{\alpha}\|^2e^{2r_+})
+
\varepsilon_-^2
\sinh^22r_-
],
\label{eqn:FQ12Opt}
\end{align}
with
\begin{equation}
\bm{\alpha}
=
\pm 
\|\bm{\alpha}\|
e^{-\frac{i}{2}(\eta+\chi+\phi)}
\begin{pmatrix}
\smallskip
1\\
0
\end{pmatrix}
\quad\text{and}\quad
\theta=0.
\label{eqn:Opt12Mix}
\end{equation}

\begin{figure*}
\begin{tabular}{l@{\qquad\quad}l}
(a)&(b)\\
\includegraphics[scale=0.78]{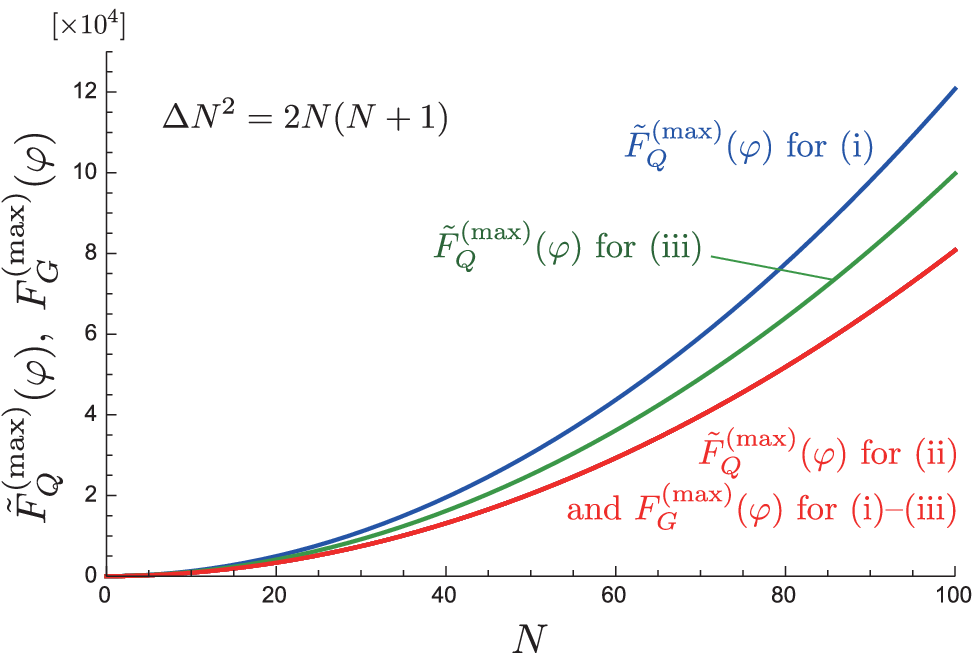}
&
\includegraphics[scale=0.78]{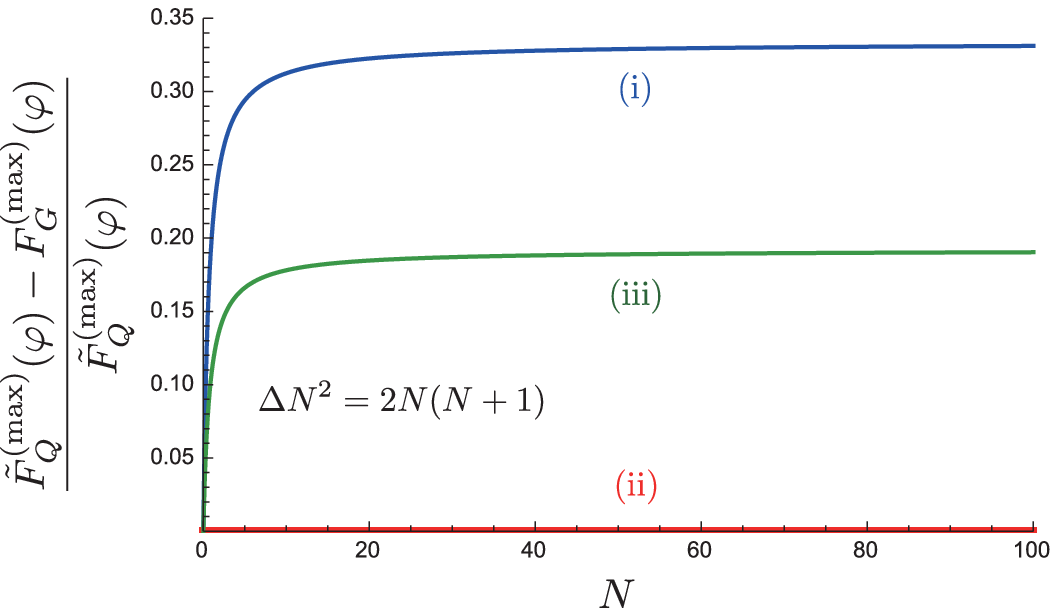}
\end{tabular}
\caption{(a) The optimal QFI $F_G^\text{(max)}(\varphi)$ in Eq.\ (\ref{eqn:FoptN}) with the Gaussian input state (\ref{eqn:OptGaussState}) compared with the {maximal value of the} QFI $F_Q^\text{(max)}(\varphi)$ in Eq.\ (\ref{eq:central}) with $\Delta N^2=2N(N+1)$, i.e., $\tilde{F}_Q^\text{(max)}(\varphi)$ in Eq.\ (\ref{eqn:FQmaxComp}),  for the three simple cases (i), (ii), and (iii) in Fig.\ \ref{fig:SpecialCases}  with $(\varepsilon_+,\varepsilon_-)=(1,-1)$, $(1,1)$, and $(1,0)$, respectively. (b) The gaps between $F_G^\text{(max)}(\varphi)$ and $\tilde{F}_Q^\text{(max)}(\varphi)$ for (i)--(iii).}
\label{fig:Fopt}
\end{figure*}
The final step is the maximization of the QFI with respect to the squeezing parameters $\{r_+,r_-\}$ and the norm $\|\bm{\alpha}\|^2$, included in the constraint (\ref{eqn:ConstraintN}).
In order to fulfill this condition, we insert
\begin{equation}
\sinh^2r_-=N-\sinh^2r_+-\|\bm{\alpha}\|^2
\label{eqn:ConstraintNr2}
\end{equation}
into Eq.\ (\ref{eqn:FQ12Opt}).
It yields
\begin{widetext}
\begin{equation}
\max_{\theta,\,\bm{n}} F_G(\varphi)
=8
\varepsilon_+^2
\sinh^2r_+
(1+\sinh^2r_+)
+4\varepsilon_+^2\|\bm{\alpha}\|^2e^{2r_+}
+8
\varepsilon_-^2
(N-\sinh^2r_+-\|\bm{\alpha}\|^2)
(1+N-\sinh^2r_+-\|\bm{\alpha}\|^2).
\end{equation}
The optimization must take into account the positivity of the right-hand side of Eq.\ (\ref{eqn:ConstraintNr2}) for a given $\|\bm{\alpha}\|^2$.
The maximum is actually reached with the largest possible $r_+$, namely, with $r_-=0$. 
We obtain
\begin{equation}
\max_{\text{given}\,\|\bm{\alpha}\|^2}
F_G(\varphi)
=4
{\varepsilon_+^2}
\,\Bigl[
2
(N-\|\bm{\alpha}\|^2)
(1+N-\|\bm{\alpha}\|^2)
+\|\bm{\alpha}\|^2
\left(
1+2N-2\|\bm{\alpha}\|^2
+2
\sqrt{
(N-\|\bm{\alpha}\|^2)
(1+N-\|\bm{\alpha}\|^2)
}
\right)
\Bigr].
\label{eqn:FoptNa}
\end{equation}
\end{widetext}
This is a monotonically decreasing function of $\|\bm{\alpha}\|^2$, and reaches its maximum at $\|\bm{\alpha}\|^2=0$, that is,
\begin{equation}
F_G^\text{(max)}(\varphi)
=8 {\varepsilon_+^2}
N(N+1).
\label{eqn:FoptN}
\end{equation}
This is the {maximal value of the} QFI reachable using a pure Gaussian state {as the input. In particular, the optimal input state is}
\begin{equation}
\ket{\psi_G}
=
{\ket{r}_+\otimes\ket{0}_-,} 
\ \ %
r=\frac{1}{2}\ln\!\left(
1+2N+2\sqrt{N(N+1)}
\right),
\label{eqn:OptGaussState}
\end{equation}
where $\ket{r}$ is a squeezed vacuum state.

Remarkably, the optimal Gaussian states in Eq.\ (\ref{eqn:OptGaussState}) are product states with no entanglement.
Moreover, it is suggested to inject all the available energy into a single mode.
The result also shows that squeezing is more effective than displacing in improving the precision of the estimation with Gaussian light (in contrast to the claim in Ref.\ \cite{ref:GaussOptBraunTrepFabre}) (note that the actual state to be injected into the two ports of the given target circuit is not simply the squeezed vacuum in one port $\hat{a}_+$ with the vacuum in the other $\hat{a}_-$, but can be generated from it by transforming it through an appropriate linear circuit representing the unitary transformation relating the physical modes $\hat{a}_\pm$ to the normal modes $\hat{c}_\pm$).

In the optimal Gaussian states (\ref{eqn:OptGaussState}), the variance of the number of probe particles $\hat{N}$ in Eq.\ (\ref{enne}) is given by
\begin{equation}
(\Delta\hat{N})_G^2
=2N(N+1).
\label{eqn:VarNGauss}
\end{equation}
Therefore, the above result $F_G^\text{(max)}(\varphi)$ in Eq.\ (\ref{eqn:FoptN}) with the optimal Gaussian state should be compared with the {maximal value of the} QFI $F_Q^\text{(max)}(\varphi)$ in Eq.\ (\ref{eq:central}) with the variance $\Delta N^2$ fixed at the value given by Eq.\ (\ref{eqn:VarNGauss}), i.e.,
\begin{align}
\tilde{F}_Q^\text{(max)}(\varphi)
=\Bigl(&
|\varepsilon_+-\varepsilon_-|\sqrt{N(3N+2)}
\nonumber\\
&{}+|\varepsilon_++\varepsilon_-|\sqrt{2N(N+1)}
\Bigr)^2.
\label{eqn:FQmaxComp}
\end{align}
As expected, in the general case   there is a gap between $F_G^\text{(max)}(\varphi)$ and $\tilde{F}_Q^\text{(max)}(\varphi)$: it increases with $N$ and
 approaches 
\begin{align}
&\frac{
\tilde{F}_Q^\text{(max)}(\varphi)
-
F_G^\text{(max)}(\varphi)
}{\tilde{F}_Q^\text{(max)}(\varphi)
}
\nonumber\\
&\qquad
\to
\frac{|\varepsilon_+-\varepsilon_-|
\,\Bigl(
|\varepsilon_+-\varepsilon_-|
+2(\sqrt{6}-2)|\varepsilon_++\varepsilon_-|
\Bigr)
}{
\Bigl(
\sqrt{3}\,|\varepsilon_+-\varepsilon_-|
+\sqrt{2}\,|\varepsilon_++\varepsilon_-|
\Bigr)^2
}
\end{align}
in the  asymptotic limit $N\to\infty$ (see Fig.\ \ref{fig:Fopt}).
We notice also that the gap closes exactly 
  when $\varepsilon_+=\varepsilon_-$:  indeed we already knew that any input states with average $N$ and variance $\Delta N^2$  achieves {the maximal} QFI $F_Q^\text{(max)}(\varphi)$ in Eq.\ (\ref{eq:central}).

\section{Conclusions}\label{sec:conclusion}
In this article we have addressed the problem of the optimization of QFI when identical bosonic particles are used as probes. 
We are in particular interested in the case where the number of probe particles is not fixed but can fluctuate.
The optimization problem can be rephrased in terms of a problem of classical probability, and we have solved it for the most generic two-mode linear circuit.
We have found a concise expression for the maximal QFI, which tells us how the fluctuation in the number of probe particles enhances the precision of the estimation, and we have characterized the best input state, which is a generalization of the NOON state, or a generalization of the two-mode Schr\"odinger-cat state, depending on the properties of the circuit.
We have also identified the best input state among pure Gaussian states, which are ubiquitously used in quantum optical metrology.
The setup is quite general,  includes the standard Mach-Zehnder interferometer and other linear optical circuits, and is relevant for various practical applications.

\begin{acknowledgements}
VG thanks L. Maccone for comments and discussions. 
This work was partially supported by the Italian National Group of Mathematical Physics (GNFM-INdAM, ``Progetto Giovani''), by PRIN 2010LLKJBX on ``Collective quantum phenomena: from strongly correlated systems to quantum simulators,'' by the EU Collaborative Project TherMiQ (Grant Agreement No.\ 618074), by the Top Global University Project from the Ministry of Education, Culture, Sports, Science and Technology (MEXT), Japan, by a Grant-in-Aid for Scientific Research (C) (No.\ 26400406) from Japan Society for the Promotion of Science (JSPS), and by a Waseda University Grant for Special Research Projects (No.\ 2015K-202).
\end{acknowledgements}

\end{document}